\documentclass[a4paper,11pt]{article}
\usepackage{template}

\usepackage[T1]{fontenc}

\usepackage{amscd, mathrsfs}
\usepackage{amsthm}
\usepackage{cases}
\usepackage{here}
\usepackage{bm}

\definecolor{Pistachio}{cmyk}{0.41, 0.27, 0.45, 0}
\definecolor{Pastelpink}{cmyk}{0.21, 0.36, 0.24, 0}

\usepackage[colorlinks=true
,urlcolor=Pistachio
,anchorcolor=Pistachio
,citecolor=Pistachio
,filecolor=Pistachio
,linkcolor=Pastelpink
,menucolor=Pastelpink
,pdfproducer=medialab
,pdfa=true
]{hyperref}

\newcommand{\pdiff}[2]{\frac{\partial#1}{\partial#2}}
\newcommand{\eq}[1]{\begin{equation}\begin{split}#1\end{split}\end{equation}}

\title{\boldmath Baryon isocurvature constraints  on the primordial hypermagnetic fields}

\author[a]{Kohei Kamada,}
\author[a,b]{Fumio Uchida,}
\author[a,b,c,d]{and Jun'ichi Yokoyama}

\affiliation[a]{Research Center for the Early Universe (RESCEU), Graduate School of Science,\\The University of Tokyo, Tokyo 113-0033, Japan}
\affiliation[b]{Department of Physics, Graduate School of Science,\\The University of Tokyo, Tokyo 113-0033, Japan}
\affiliation[c]{Kavli Institute for the Physics and Mathematics of the Universe (Kavli IPMU),\\WPI, UTIAS, The University of Tokyo, Kashiwa, Chiba 277-8568, Japan}
\affiliation[d]{Trans-scale Quantum Science Institute,\\The University of Tokyo, Tokyo 113-0033, Japan}

\emailAdd{kohei.kamada@resceu.s.u-tokyo.ac.jp}
\emailAdd{fuchida@resceu.s.u-tokyo.ac.jp}
\emailAdd{yokoyama@resceu.s.u-tokyo.ac.jp}

\abstract{
It has been pointed out that hypermagnetic helicity decay at the electroweak symmetry breaking may have produced the observed baryon asymmetry of the Universe through the chiral anomaly in the standard model of particle physics.  Although fully helical magnetic field that can adequately produce the observed baryon asymmetry is not strong enough to explain the origin of the intergalactic magnetic field inferred by the Fermi satellite, the mixture of helical and nonhelical primordial magnetic fields may explain both baryogenesis and the intergalactic magnetic fields simultaneously.  We first show that such a scenario is ruled out by the constraint on the amplitude of baryon isocurvature perturbations produced by the primordial magnetic fields to avoid overproduction of deuterium at the big bang nucleosynthesis.  Then we show that any attempt to explain the origin of intergalactic magnetic field by primordial magnetogenesis before the electroweak symmetry breaking does not work due to the above constraint irrespective of the helicity and baryogenesis mechanism.

}

\begin{document}
\maketitle
\flushbottom

\section{Introduction}
\label{sec:intro}
The excess of baryons over anti-baryons in the present universe, or the baryon asymmetry of the universe (BAU), is a key ingredient in modern cosmology and essential for our very existence. 
It is quantified by the baryon-to-entropy ratio,
precisely determined by the observation of the cosmic microwave background (CMB)~\cite{Aghanim:2018eyx}
as
$\eta_{B} \equiv n_{B}/s=(8.718\pm0.054)\times 10^{-11}$, 
which is in good agreement with the one required for the success of the Big Bang Nucleosynthesis (BBN)~\cite{Fields:2019pfx}. 
The Standard Model of particle physics (SM) is hard to explain this tiny but non-zero 
asymmetry, especially in the context of inflationary cosmology, 
and its origin has been regarded as 
a fundamental problem in particle physics and cosmology.
A dynamical mechanism, baryogenesis, thus should have taken place in the early Universe
after the end of inflation before the BBN, since inflation dilutes
away the preexisting asymmetry. 

To generate non-vanishing baryon asymmetry
we must satisfy 
the Sakharov criteria~\cite{Sakharov:1967dj}:
baryon-number violation, C and CP violation, 
and departure from thermal equilibrium. 
Since they are hardly satisfied in the realms of the SM, 
new physical degrees of freedom beyond the SM are often introduced, {\it e.g.}, 
new fields to make the electroweak symmetry breaking (EWSB) strongly first order 
and to give the source of CP violation for the electroweak baryogenesis~\cite{Kuzmin:1985mm}, 
or the right-handed neutrinos for leptogenesis~\cite{Fukugita:1986hr}. 
In this regard, 
a baryogenesis scenario from hypermagnetic helicity decay~\cite{Giovannini:1997gp,Giovannini:1997eg, Bamba:2006km, Fujita:2016igl,Kamada:2016eeb, Kamada:2016cnb} is remarkable since it 
does not require any new degrees of freedom beyond the SM in the mechanism itself. 
It utilizes 
the chiral anomaly in the SM \cite{Adler:1969gk, Bell:1969ts, tHooft:1976rip}, 
which relates the changes in the global baryon (B) and lepton (L) number ($Q_B$ and $Q_L$) 
to the changes in the SU(2)$_{L}$ Chern-Simons number ($N_\mathrm{CS}^{L}$) and the U(1)$_{Y}$ hypermagnetic helicity ($\mathcal{H}_{Y}$), 
\eq{\label{eq:ABJ}
    \Delta Q_B = \Delta Q_L = N_{\rm g} \left(\Delta N_\mathrm{CS}^{L}- \frac{g'^2}{16 \pi^2} \Delta \mathcal{H}_{Y}\right), 
    }
where  $N_{\mathrm{g}}=3$ is the number of fermion generations and $g'$ is the gauge coupling for the U(1)$_{Y}$ hypergauge interaction. 
Thus baryon asymmetry can be generated if the hypermagnetic helicity decay occurs. 
This idea can be realized in realistic cosmological scenarios such as 
the axion inflation~\cite{Turner:1987bw,Garretson:1992vt,Anber:2006xt}, the SU(5) Grand Unified Theory baryogenesis~\cite{Yoshimura:1978ex,Ignatiev:1978uf,Weinberg:1979bt} 
through the chiral plasma instability~\cite{Joyce:1997uy,Tashiro:2012mf,Akamatsu:2013pjd}, 
or the Affleck-Dine mechanism~\cite{Affleck:1984fy,Dine:1995kz}, 
in which the long-range U(1)$_Y$ hypermagnetic fields are generated before the EWSB with helicity  (See, {\it e.g.}, Refs.~\cite{Anber:2015yca,Adshead:2016iae,Jimenez:2017cdr,Domcke:2018eki,Domcke:2019mnd,Barrie:2020kpt,Kamada:2018tcs,Kamada:2019uxp}). 
The hypermagnetic helicity decay occurs mainly at the EWSB when the hypermagnetic fields are converted into the U(1)$_{\mathrm{EM}}$ 
magnetic fields. 
The Sakharov criteria are satisfied by the SM chiral anomaly (B-violation) and 
the existence of the long-range helical hypermagnetic fields (C and CP-violation as well as the deviation from 
thermal equilibrium). 
It has been shown that in the electroweak crossover suggested by the 125 GeV Higgs~\cite{DOnofrio:2014rug,DOnofrio:2015gop}
baryon asymmetry can remain until today~\cite{Kamada:2016cnb} against the washout by the electroweak sphalerons~\cite{Kuzmin:1985mm}.
In this scenario, the physics beyond the SM is required for magnetogenesis but not for baryogenesis. 

An advantage of this scenario is that the U(1)$_Y$ hypermagnetic fields 
that are responsible for baryogenesis can 
persist until the present time after the conversion into the U(1)$_{\mathrm{EM}}$ magnetic fields at the EWSB 
as the intergalactic magnetic fields (IGMFs), which would act as its potential probe. 
Indeed, the existence of IGMFs has been suggested by the recent TeV blazar observations 
with {\it Fermi-LAT} \cite{Neronov:1900zz, Tavecchio:2010mk, Ando:2010rb, Dolag:2010ni, Essey:2010nd, Taylor:2011bn,Takahashi:2013lba, Finke:2015ona, Biteau:2018tmv,AlvesBatista:2020oio}. The lack of secondary GeV photon in the blazar spectra, expected to be generated during the propagation in 
the intergalactic void, can be attributed to the IGMFs so that the lower bound of their strength 
and coherence length are given.
If further observation of the IGMFs can pin down their strength and coherence length
as well as their helicity, in principle we can test the scenario, 
which gives us insights into the origin of the IGMFs as well as that of BAU. 

Previous studies have already shown that maximally helical U(1)$_{Y}$ gauge fields that can explain the observationally suggested IGMF would 
suffer from baryon overproduction~\cite{Kamada:2016cnb}, 
unless the real transition from the hypermagnetic fields to the electromagnetic fields at the electroweak crossover is much more rapid than
the one extrapolated from the result of the lattice calculation~\cite{DOnofrio:2014rug,DOnofrio:2015gop}. 
One way to accommodate the baryogenesis and IGMFs indicated by the blazar observations
is to take the U(1)$_Y$ hypermagnetic fields to be {\it partially} helical, 
which might be realized by, {\it e.g.}, an incomplete cancellation of the hypermagnetic helicity and 
the chiral asymmetry carried by the right-handed electrons generated 
in axion inflation via the chiral plasma instability~\cite{Domcke:2019mnd}.\footnote{Another option 
is to consider the second magnetogenesis after the EWSB. In this case we do not have direct
relationship between the BAU and IGMFs.}
In such a case, the baryogenesis is less effective so that stronger hyoermagnetic fields,
possibly up to the ones consistent with the lower bound of the IGMFs, 
are needed to explain the present BAU. 
The purpose of this paper is to address further this option. 

The key ingredient to investigate this situation
but not to have been explored in depth yet 
is the baryon isocurvature perturbation, 
which should be generated at the scale corresponding to the coherence length of the hypermagnetic fields. 
It is well known that the baryon isocurvature perturbations are constrained 
by the observations of cosmic microwave background (CMB) on large scales ($k\lesssim 0.1 \;\mathrm{Mpc}^{-1}$; $k$ is the comoving wavenumber)~\cite{Akrami:2018odb}, 
which would give constraints on acausally generated hypermagnetic fields that never entered 
the hydrodynamic turbulence due to sufficiently large coherence length.\footnote{The baryon isocurvature perturbations are degenerate with cold dark matter isocurvature perturbations. 
It has been claimed that 21 cm fluctuations can potentially discriminate the degeneracy \cite{Kawasaki:2011ze},
but still it is difficult to distinguish tiny isocurvature fluctuations \cite{Takeuchi:2013hza}.} 
The hypermagnetic fields with smaller coherence length, {\it e.g.}, the ones that have once entered the hydrodynamic turbulence regime, 
do not receive stringent constraints from such cosmological observations. 
However, such small-scale baryon isocurvature perturbations may spoil the successful BBN and their spectrum
can be constrained by the studies of  inhomogeneous BBN~\cite{Applegate:1987hm,Alcock:1987tx,Inomata:2018htm}. 
In particular, it has recently been pointed out that the baryon isocurvature perturbation at a scale 
larger than the neutron diffusion scale at the BBN epoch, $k\lesssim 4\times 10^8 \;\mathrm{Mpc}^{-1}$, 
is constrained by the deuterium overproduction due to the second-order effect~\cite{Inomata:2018htm}. 
As a result, one can obtain an upper bound on their coherence length.
We find that since the baryon isocurvature perturbation is generated even from non-helical hypermagnetic fields~\cite{Giovannini:1997gp,Giovannini:1997eg}, 
the constraint becomes severer for less helical hypermagnetic fields, in which case hypermagnetic fields with longer coherence length 
are needed for the BAU. 
Consequently, together with the constraint on their properties
from the magnetohydrodynamics (MHD)~\cite{Banerjee:2004df}, we find a lower bound of 
the helicity fraction, $\epsilon$, which gives an upper bound of the hypermagnetic field strength 
that can generate the present BAU. 
In terms of the present IGMF properties, the corresponding upper bound lies below the most stringent lower bound of the IGMF strength~\cite{Biteau:2018tmv}, independent of the detail of their properties. 
Moreover, since the non-helical part of the hypermagnetic fields contributes to the baryon isocurvature perturbation, 
it constrains their properties even in the case the homogeneous part of the BAU is generated by another mechanism. 
It turns out that  regardless of their helicity properties, any hypermagnetic fields with too large strength and coherence length
 are not allowed before the EWSB. 
We conclude that the IGMFs suggested by the observation of the {\it Fermi-LAT} 
collaboration must have been generated after the EWSB so as not to undermine
the success of the BBN, 
irrespective of their  helical property as well as the generation mechanism of the present BAU. 
The baryogenesis from the hypermagnetic helicity decay can still be responsible for the present BAU, 
but we need additional magnetogenesis or an unknown mechanism of the magnetic field amplification after the EWSB.

The paper is organized as follows. 
In \S \ref{sec:General treatment}, 
we briefly review  baryogenesis from  hypermagnetic helicity decay and derive the spectrum of baryon isocurvature perturbation.
We provide a general treatment with which we can assess whether this scenario survives against the baryon isocurvature constraints. 
In \S \ref{sec:AfterBG}, 
we present the general formalism to constrain the scenario from baryon isocurvature perturbation
and the way to connect the constraint to the IGMF observations,
by taking into account the evolution of the magnetic fields after the EWSB. 
In \S \ref{sec:Concrete models},
by adopting simple forms of the magnetic field power spectrum, 
we obtain  constraints on the magnetic fields  quantitatively in the case  they are responsible for the present BAU.
In \S \ref{sec:Application},
we present the constraints on the non-helical hypermagnetic fields without specifying baryogenesis models.
Finally \S \ref{sec:Discussion} is devoted to conclusion and discussion.

\section{Baryogenesis from hypermangetic helicity decay}
\label{sec:General treatment}
We begin by reviewing the baryogenesis scenario from the hypermagnetic helicity decay. 
In the previous studies~\cite{Fujita:2016igl,Kamada:2016eeb,Kamada:2016cnb} the hypermagnetic fields are parameterized by 
their characteristic field strength $B_\mathrm{p}$ and coherence length $\xi_\mathrm{p}$, assuming a delta-function-like 
spectrum. 
For our purpose to investigate the baryon isocurvature perturbation, 
we here reformulate the consequence of this scenario by taking into account the spectrum of the hypermagnetic fields, 
as is also done in Ref.~\cite{Giovannini:1997gp,Giovannini:1997eg}.

\subsection{Generation of the net baryon asymmetry}
\label{sec:Baryon asymmetry production}
Let us first give the explanation how the net baryon asymmetry is generated by the decaying hypermagnetic helicity in a realistic cosmic history 
following Ref.~\cite{Kamada:2016cnb}. 
As explained in the introduction, the change in the comoving hypermagnetic helicity density induces 
the change in the BAU through the chiral anomaly of the SM (Eq.~\eqref{eq:ABJ}). 
Here we consider the case where the hypermagnetic fields are generated before the EWSB with sufficiently large coherence length and non-vanishing net helicity. 
Then the hypermagnetic fields evolve according to MHD. 
Basically, the net hypermagnetic helicity is almost conserved with the help of large electric conductivity~\cite{Baym:1997gq,Arnold:2000dr} until the EWSB. 
However, the hypermagnetic helicity decay occurs in two ways in the cosmic history. 
The first is that the magnetic field diffusion due to the large but finite electric conductivity~\cite{Giovannini:1997gp,Giovannini:1997eg}, 
and the other is the electroweak symmetry breaking where the hypermagnetic helicity is converted
into the U(1)$_\mathrm{EM}$ magnetic helicity~\cite{Kamada:2016cnb}. 
It turned out that the latter gives more significant contributions on the BAU, 
unless the dynamics of the EWSB is significantly different from the results of the lattice simulations~\cite{DOnofrio:2014rug,DOnofrio:2015gop}
as well as the one-loop analytic estimate~\cite{Kajantie:1996qd}.
Note that after the completion of the EWSB U(1) baryon symmetry is no longer anomalous 
and baryon asymmetry is not generated from the U(1)$_\mathrm{EM}$ magnetic helicity decay. 
Below, we explain in more depth how the BAU is generated and 
how we can evaluate it. 
We here assume that the possible cancellation of the chiral asymmetry 
and hypermagnetic helicity~\cite{Domcke:2019mnd} does not occur 
and the hypermagnetic helicity is much larger than the chiral asymmetry in the system 
before the EWSB. 

To evaluate the BAU generated in this mechanism, we need to take into account the washout effects
by the electroweak sphalerons~\cite{Kuzmin:1985mm}, 
since this mechanism generates only $B+L$ but not $B-L$ asymmetry. 
If the EWSB would have completed earlier than the freezeout of the electroweak sphalerons, 
the baryon asymmetry would not remain much (See, however, Ref.~\cite{Kamada:2016eeb}). 
However, lattice simulations of the EWSB in the light of the 125 GeV Higgs suggest that 
they occur at almost the same time so that the electroweak sphaleron washout is incomplete. 
Then the evolution of the mean baryon asymmetry can be evaluated by solving a simplified 
kinetic equation for the mean baryon-to-entropy ratio, $\overline{\eta}_B$~\cite{Kamada:2016cnb}, 
\eq{\label{eq:KineticEqn}
    \frac{d\overline{\eta}_B}{dx} 
    = (\text{source}) - (\text{washout}), 
    }
with the dominant contributions around the EWSB ($T\simeq 130 - 145$\footnote{Electron Yukawa is dominant for the washout at $T \gtrsim 145$ GeV.} GeV) being~\cite{Kamada:2016cnb}
\eq{
    (\text{source})=
    \frac{3}{2}(g^2+g'^2)\frac{d\theta_{\mathrm{w}}}{d\ln x} \sin 2 \theta_\mathrm{w} \mathcal{S}_{\mathrm{AB}},
    \quad (\text{washout})=
    \frac{111}{34}\gamma_{\mathrm{w, sph}}\overline{\eta}_B,
    }
where 
\eq{\label{eq:source}
    \mathcal{S}_{\mathrm{AB}}
    =\frac{H}{8\pi^2a^3sT}{\overline{\boldsymbol{A}_{\mathcal{A}}\cdot\boldsymbol{B}_{\mathcal{A}}}}.
    }
with the overline representing the volume average. 
Here $x = M_0/T$ with $M_0 \equiv M_\mathrm{pl}/\sqrt{\pi^2  g_*/90}$  
($M_\mathrm{pl}\simeq 2.43 \times 10^{18}$ GeV is the reduced Planck mass
and $g_*$ is the effective number of relativistic degrees of freedom during the process.)
and $T$ being the temperature of the Universe.
$g$ is the gauge coupling for the SU(2)$_L$ electroweak gauge interaction 
and $\theta_\mathrm{w}$ is the temperature-dependent effective weak mixing angle, 
which gradually changes from 0 to $\theta_{\mathrm{w}0} \equiv \tan^{-1} (g'/g)$ at the electroweak crossover.
$\gamma_{\mathrm{w, sph}}\simeq \exp[-(146.6\pm 1.0)+(0.83\pm0.01) T/\mathrm{GeV}]$~\cite{DOnofrio:2014rug} 
is the transport coefficient for the electroweak sphalerons.
$H\equiv {\dot a}/a$ is the Hubble parameter with $a$ being the scale factor and $s=(2 \pi^2/45)g_{*s}T^3$ is the entropy density
($g_{*s}$ is the relativistic degrees of freedom for entropy). 
Here ${\cal A}$ stands for the massless U(1) gauge field during the electroweak crossover, 
and ${\bm A}_{\cal A}$ and ${\bm B}_{\cal A}$ are the 
vector potential and comoving magnetic fields for ${\cal A}$, respectively. 
They are characterized by the effective weak mixing angle $\theta_\mathrm{w}$ as
${\bm Y} = \cos \theta_\mathrm{w} {\bm A}_{\cal A}$ and ${\bm W}^3 = \sin \theta_\mathrm{w} {\bm A}_{\cal A}$, 
with ${\bm Y}$ and ${\bm W}^3$ being the vector potentials for the U(1)$_Y$ hyper gauge interaction 
and the third component of the SU(2)$_L$ electroweak gauge interaction. 
Here the kinetic equation is derived by taking into account the equilibrium conditions 
of the rapid spectator processes such as the spin-flipping process of chiral fermions through 
the Yukawa interactions other than the one for the electrons and the strong sphalerons. 
Note that during the electroweak crossover, the long-range ``magnetic'' field turns from that of U(1)$_Y$ 
hypergauge interaction to that of the  U(1)$_\mathrm{EW}$ electromagnetic interaction 
through the massless U(1) gauge field ${\cal A}$. 
Thus the right hand side of  the anomaly equation, $(g^2/16\pi^2)\mathrm{Tr}\left[W_{\mu\nu}\Tilde{W}^{\mu\nu}\right]-(g'^2/32\pi^2)Y_{\mu\nu}\Tilde{Y}^{\mu\nu}$, 
includes the terms, {\it e.g.},  ${d ( \sin \theta_\mathrm{w} {\bm A}_{\cal A}})/d \ln x \cdot {\bm B}_{\cal A} \sim ( d  \theta_\mathrm{w}/d \ln x) {\bm A}_{\cal A} \cdot{\bm B}_{\cal A}$.\footnote{We carry out all the computation with
the Coulomb gauge 
${\bm \nabla} \cdot {\bm A}_{\cal A} =0$, but the final results are gauge independent~\cite{Kamada:2016cnb}}.
The gradual changes of the weak mixing angle during the electroweak crossover
give significant contributions to the source term. 
Around the electroweak symmetry breaking, the electroweak sphaleron becomes less and less effective 
and turns to the rate-determining reaction. 

When the transport coefficient of the electroweak sphalerons is larger than the Hubble rate, 
the system enters the equilibrium state with $d\overline{\eta}_B/dx = 0$  for Eq.~\eqref{eq:KineticEqn}, 
which gives the mean baryon asymmetry as 
\begin{equation}
    \overline{\eta}_B \simeq \frac{17}{37} \frac{g^2+g'^2}{\gamma_{\mathrm{w, sph}}} \frac{d\theta_{\mathrm{w}}}{d\ln x} \sin 2 \theta_\mathrm{w} \mathcal{S}_{\mathrm{AB}}, \label{equil}
\end{equation}
before the electroweak sphaleron freezeout. 
As the temperature of the Universe decreases, the electroweak sphaleron freezes out around $T=T_\mathrm{fo} \simeq 135$ GeV. 
(Hereafter the subscript fo represents that the quantity is evaluated at the sphaleron freeze out, $T=T_\mathrm{fo}$.)
Just after the sphaleron freezeout the electroweak symmetry breaking 
completes and there will be no longer significant induction of baryon asymmetry. 
Numerically it is found that the resultant baryon asymmetry is evaluated the equilibrium solution~\eqref{equil}
at $T_{\mathrm{fo}} \simeq 135$ GeV~\cite{Kamada:2016cnb}. 
Since at the electroweak sphaleron freezeout, the transport coefficient of the electroweak sphaleron is 
related to the Hubble parameter as 
$\gamma_{\mathrm{w, sph}}(T_{\mathrm{fo}}) \simeq H(T_{\mathrm{fo}})/T_{\mathrm{fo}}$, 
the resultant baryon asymmetry is evaluated in a simplified form as 
\eq{\label{eq:ProducedBAU}
    \overline{\eta}_B
    =\left.\frac{17}{296\pi^2}(g^2+g'^2)\frac{d \theta_\mathrm{w}}{d \ln x} \sin 2 \theta_\mathrm{w}\frac{\overline{\boldsymbol{A}_{\mathcal{A}}\cdot\boldsymbol{B}_{\mathcal{A}}}}{a^3s}\right|_{T=T_\mathrm{fo}}.
    }
Therefore, if we have non-vanishing $\overline{\boldsymbol{A}_{\mathcal{A}}\cdot\boldsymbol{B}_{\mathcal{A}}}$ 
on average, we will obtain non-vanishing net baryon asymmetry. 
For a peaky spectrum of the hypermagnetic fields with the characteristic comoving field strength $B_\mathrm{c}$ 
(or the physical strength $B_\mathrm{p} = a^{-2} B_\mathrm{c}$) 
and the comoving coherence length $\xi_\mathrm{c}$ (or the physical coherence length $\xi_\mathrm{p} = a \xi_\mathrm{c}$)
with the helicity fraction $\epsilon$, we can write  $\overline{\boldsymbol{A}_{\mathcal{A}}\cdot\boldsymbol{B}_{\mathcal{A}}} \simeq \epsilon \xi_\mathrm{c} B_\mathrm{c}^2 = a^3 \epsilon \xi_\mathrm{p} B_\mathrm{p}^2$ and evaluate the net baryon asymmetry as 
\begin{align}\label{eq:ProducedBAUBlambda}
    \overline{\eta}_B
    &=\left.\frac{17}{296\pi^2}(g^2+g'^2)\frac{d \theta_\mathrm{w}}{d \ln x} \sin 2 \theta_\mathrm{w}\frac{\epsilon \xi_\mathrm{p} B_\mathrm{p}^2}{s}\right|_{T=T_\mathrm{fo}} \notag \\
    & \sim 10^{-10} \epsilon \left.\sin 2 \theta_\mathrm{w} \frac{d \theta_\mathrm{w}}{d \ln x} \left(\frac{\xi_\mathrm{p} }{10^6\mathrm{GeV}^{-1}}\right) \left(\frac{B_\mathrm{p}}{10^{-3} \mathrm{GeV}^2}\right)^2\right|_{T=T_\mathrm{fo}}, 
\end{align}
which will be more formally derived in the next subsection in terms of the magnetic field power spectrum. 

The value of $\sin 2 \theta_\mathrm{w} d \theta_\mathrm{w}/d \ln x$ at $T=T_\mathrm{fo}$ has 
relatively large uncertainties in the analytic expressions~\cite{Kajantie:1996qd} and results of the lattice calculations~\cite{DOnofrio:2015gop}. 
On the one hand, the time dependence of the weak mixing angle is evaluated analytically at the one-loop level as~\cite{Kajantie:1996qd}
\eq{
    \cos^2\theta_{\rm w}=\cos^2\theta_{{\rm w}0}\left(1+\frac{11}{12}\frac{g_3^2\sin^2\theta_{{\rm w}0}}{\pi m_W(T)}\right),
    }
where $\theta_{{\rm w}0}$ is the weak mixing angle at zero temperature,
$g_3$ is the three dimensional SU($2$) gauge coupling,
and $m_W(T)$ is the perturbative $W$ boson mass.
By substituting
\eq{
    g_3^2\simeq g^2 T,\quad m_W(T)=\frac{g\phi(T)}{2},
    }
with adopting the fitting function to the numerical lattice result for the Higgs condensate, $\phi(T)$, around the EWSB, $T\lesssim162$ GeV \cite{Kamada:2016eeb},
\eq{
    \phi(T)\simeq0.23T\sqrt{162-\frac{T}{1\;{\rm GeV}}},
    }
we obtain
\eq{\label{eq:1-loop}
    \left.\frac{d \theta_\mathrm{w}}{d \ln x} \sin 2 \theta_\mathrm{w}\right|_{T=T_\mathrm{fo}}=
    \left.-T\frac{d \theta_\mathrm{w}}{d T} \sin 2 \theta_\mathrm{w}\right|_{T=T_\mathrm{fo}}
    \simeq 0.14,
    }
where we have used $g=0.65$ and $\cos^2\theta_{{\rm w}0}=0.77.$

On the other hand, the three dimensional lattice calculation of the weak mixing angle has been calculated in the light of the $125$ GeV Higgs, for $140$ GeV $<T< 170$ GeV in Ref. \cite{DOnofrio:2015gop}.
In principle the lattice calculation is an all-orders calculation that includes even the non-perturbative effects.
However, the result has relatively large errors and agrees with the analytic calculation only marginally.
Since one cannot tell which estimate is more reliable and the analytic estimate is applicable only limited temperature ranges,
which is not suitable for solving the kinetic equation numerically,
Ref. \cite{Kamada:2016cnb} has adopted a phenomenological fitting formula with a smoothed step function for $\theta_{\rm w}$ as
\eq{\label{eq:EWSB_KLpara}
    \cos^2\theta_{\mathrm{w}}(T)
    =\cos^2\theta_{\mathrm{w}0}+\frac{1-\cos^2\theta_{\mathrm{w}0}}{2}\left(1+\tanh\frac{T-T_{\mathrm{step}}}{\Delta T}\right).
    }
The values of the parameters $T_{\mathrm{step}}$ and $\Delta T$, chosen in Ref. \cite{Kamada:2016cnb}, which give relatively good fit for the lattice results are shown in Table \ref{tb:EWSB}.\footnote{The parameterization A in Ref.~\cite{Kamada:2016cnb} is omitted since it does not give a good fit at $T<160$ GeV and Eq.~\eqref{eq:ProducedBAUBlambda} is not applicable for the estimate of the resultant baryon asymmetry~\cite{Kamada:2016cnb}.}
\begin{table}
\begin{center}
  \begin{tabular}{c||ccccc|c}
    parameterizations & B & C & D & E & one-loop analytic\\ \hline\hline
    $T_{\mathrm{step}}/\mathrm{GeV}$  & $160$ & $160$ & $155$ & $155$ & -\\
    $\Delta T/\mathrm{GeV}$  & $5$ & $10$ & $10$ & $20$ & -\\ \hline
    $\left.-T\dfrac{d \theta_\mathrm{w}}{d T} \sin 2 \theta_\mathrm{w}\right|_{T=T_\mathrm{fo}}$  & $6\times10^{-4}$ & $4\times10^{-2}$ & $0.1$ & $0.3$ & $0.14$
  \end{tabular}
  \caption{Parameters that characterize the EWSB are shown. Parameterizations B-E are the ones with Eq.~\eqref{eq:EWSB_KLpara} chosen in Ref.~\cite{Kamada:2016cnb}, and the last column is the analytic estimate (Eq.~\eqref{eq:1-loop}) with the the formula at the one-loop level 
in Ref.~\cite{Kajantie:1996qd}. We use the latter as the fiducial value of the temperature dependence of the weak mixing angle 
for the estimate of the resultant baryon asymmetry.}
  \label{tb:EWSB}
\end{center}
\end{table}
Once more, we do not yet have a definite answer for the estimate of the precise time dependence of the weak mixing angle and we admit typically $\mathcal{O}(10^3)$ uncertainty in the estimate of $d\theta_{\rm w}/dT.$

For the practical purpose, however, we adopt Eq. \eqref{eq:1-loop} as the fiducial value for the estimate of the baryon asymmetry,
since the perturbative analytic estimate becomes more accurate at lower temperature $T<140$ GeV,
whereas the lattice calculation gives larger errors for lower temperature\footnote{We are grateful to Mikko Laine for pointing it out.}.
Combining Eqs.~\eqref{eq:ProducedBAUBlambda} and~\eqref{eq:1-loop} (and even using Tab.~\ref{tb:EWSB}), we can see that it is possible to explain the present BAU with appropriate properties of primordial hypermagnetic fields.

\subsection{Magnetic field spectrum}
\label{sec:Definition}
To investigate the spatial distribution of the baryon asymmetry, 
it is clear from the right hand side of Eq.~\eqref{eq:ProducedBAU}
that we need to know the spatial distribution of the magnetic fields. 
In this subsection, we provide the way to parameterize the spatial distributions of magnetic fields, namely, 
their power spectrum, by assuming that they act as stochastic fields, 
which is often realized in cosmic history from some magnetogenesis mechanisms 
such as those from axion inflation~\cite{Turner:1987bw, Garretson:1992vt,Anber:2006xt}, standard inflation with a dilatonic coupling in the kinetic function of gauge fields \cite{Ratra:1991bn,Bamba:2003av,Martin:2007ue}, or the strong first order 
phase transitions~\cite{Hogan:1983zz,Quashnock:1988vs,Vachaspati:1991nm}. 
In the next subsection, we provide the expressions of the baryon isocurvature perturbations
in terms of the magnetic field power spectrum provided in this subsection. 
Here we consider that the magnetic field power spectrum smoothly converts from the one for the U(1)$_Y$ hypergauge field
and to the one for the U(1)$_\mathrm{EM}$ electromagnetic field without significant decay or change of the scales
since the change of the weak mixing angle is a relatively slow process during the electroweak crossover~\cite{DOnofrio:2015gop}. 
In what follows we do not distinguish the hypergauge field and the electromagnetic field, as well as the intermediate
massless gauge field ${\cal A}$ and simply denote them as $A_{\mu}$ as long as the power spectrum is concerned.

Let us now study the distribution of the gauge field $A_i$ in the wavenumber space\footnote{We adopt $f(k) = \int dx e^{-ikx} f(x)$ and $f(x) = \int dk/(2\pi) e^{ikx} f(k)$ as the Fourier transformation.}
by adopting the Coulomb gauge (See the Appendix \ref{sec:Gauge choice} for more explanation on the gauge fixing). 
The power spectrum of the vector potential is then defined as follows,
\eq{
    \langle A^*_i(\boldsymbol{k},t)A_j(\boldsymbol{k}',t)\rangle
    =(2\pi)^3\delta^3(\boldsymbol{k}-\boldsymbol{k}')\mathcal{F}^{A}_{ij}(\boldsymbol{k},t),
    }
with
\eq{\label{eq:kspaceCorr}
    \mathcal{F}^{A}_{ij}(\boldsymbol{k},t)
    =P_{ij}(\hat{\boldsymbol{k}})S(k,t)+i\epsilon_{ijm}\hat{k}_mA(k,t),
    }
where $\epsilon_{ijm}$ is the 3-dimensional Levi-Civita tensor with $\epsilon_{123}=1$, and
\begin{equation}
    P_{ij}(\hat{\boldsymbol{k}}) \equiv \delta_{ij}-\hat{k}_i\hat{k}_j, \quad {\hat k}_i \equiv \frac{k_i}{k}, \quad k \equiv |{\bm k}|. 
\end{equation}
$S(k,t)$ and $A(k,t)$ are the symmetric and anti-symmetric parts of the power spectrum of the magnetic fields, respectively. 
We keep them arbitrary for the moment to make our discussion general. 
Hereafter we omit the argument $t$
but implicitly assume the time dependence of these functions, which should obey the MHD.
Assuming the Gaussianity of the magnetic field stochastic distribution,
we can characterize all the properties of the magnetic fields solely by these two functions, $S(k)$ and $A(k)$,
since the higher-order cumulants vanish for Gaussian-distributed stochastic variables.
The comoving energy density, helicity density, and coherence length of the magnetic fields are now formally defined as
\begin{align}
    \mathcal{E}_\mathrm{c}
    &=\frac{1}{2V} \int d^3 x \langle {\bm B}({\bm x})^2 \rangle \notag\\
    &= \frac{1}{2} \int \frac{d^3k}{(2\pi)^3}k^2\sum_i \mathcal{F}^{A}_{ii}(\boldsymbol{k})=\frac{1}{2\pi^2}\int dk k^4
    S(k), \label{eq:Strength}\\
    h_\mathrm{c}
    &= \frac{1}{V} \int d^3 x \langle {\bm A} ({\bm x}) \cdot {\bm B}({\bm x})\rangle \notag\\
    &=  -i\epsilon_{ijm}\int \frac{d^3k}{(2\pi)^3}k_m\mathcal{F}^{A}_{ij}(\boldsymbol{k})
    =\frac{1}{\pi^2}\int dk k^3 A(k), \label{eq:Helicity}\\ 
    \xi_\mathrm{c}
    &=\frac{\int dk k^3 S(k)}{\int dk k^4 S (k)}, \label{eq:CLength}
\end{align}
respectively, where $V = \int d^3 x$ is the volume factor. 
The helicity fraction $\epsilon$, which is in principle time-dependent,  is defined as 
\begin{equation}
    \epsilon \equiv \frac{h_\mathrm{c}}{2 \xi_\mathrm{c} {\cal E}_\mathrm{c}}. 
\end{equation}
Now from Eq.~\eqref{eq:ProducedBAU} the net mean baryon asymmetry is given in terms of the power spectrum as
\begin{align} \label{eq:ProducedBAU_xiER}
    {\overline \eta_B} & = \left.  \frac{17}{296 \pi^2} \frac{g^2+g'^2}{a^3 s} \frac{d \theta_\mathrm{w}}{d \ln x} \sin 2 \theta_\mathrm{w} \frac{1}{ \pi^2 } \int dk k^3 A(k)\right|_{T=T_\mathrm{fo}} \notag \\
    &\equiv  \left. {\cal C} h_\mathrm{c} \right|_{T=T_\mathrm{fo}} = \left. 2{\cal C} \epsilon \xi_\mathrm{c} {\cal E}_\mathrm{c} \right|_{T=T_\mathrm{fo}},  
\end{align}
where we have defined 
\begin{align}\label{eq:defcalC}
    {\cal C} & \equiv  \left.  \frac{17}{296 \pi^2} \frac{g^2+g'^2}{a^3 s} \frac{d \theta_\mathrm{w}}{d \ln x} \sin 2 \theta_\mathrm{w}\right|_{T=T_\mathrm{fo}} \notag \\
    &\simeq 2 \times 10^{34} \mathrm{GeV}^{-3}\left(\dfrac{\left.\frac{d \theta_\mathrm{w}}{d \ln x} \sin 2 \theta_\mathrm{w}\right|_{T=T_\mathrm{fo}}}{0.14}\right)  \simeq  1 \times 10^{33} \;\mathrm{Mpc}^{-1} {\rm G}^{-2}\left(\dfrac{\left.\frac{d \theta_\mathrm{w}}{d \ln x} \sin 2 \theta_\mathrm{w}\right|_{T=T_\mathrm{fo}}}{0.14}\right).
\end{align}
Here we have used $g'=0.35, g=0.65, g_{*s}^\mathrm{fo} = 106.75$, and $a_\mathrm{fo} = 5.8 \times 10^{-16}$ (for  $T_\mathrm{fo}=135$ GeV). 
We have adopted the natural Heaviside-Lorentz units,\footnote{In this unit system, $\hbar = c = \varepsilon_0 = 1$, while the natural Gaussian CGS units set $\hbar = c = 4\pi\varepsilon_0 = 1$.}
whence $1\,\textrm{G} =1.95 \times 10^{-20}\,\textrm{GeV}^2$ and $1\,\textrm{Mpc} = 1.56\times 10^{38}\,\textrm{GeV}^{-1}$.

In the case when $S(k)$ is proportional to $A(k)$, 
the helicity fraction is found to be the proportional constant, 
$A(k)=\epsilon S(k)$. 
Note that the realizability condition~\cite{1978mfge.book.....M}  imposes $|\epsilon|\leq1$.
In particular, by requiring $\overline{\eta}_B \simeq 10^{-10}$, the following
relation applies.
\eq{\label{eq:BAUEWSB}
  \left(\dfrac{\left.\frac{d \theta_\mathrm{w}}{d \ln x} \sin 2 \theta_\mathrm{w}\right|_{T=T_\mathrm{fo}}}{0.14}\right)^{\frac{1}{2}}  \left(\frac{\xi_{\rm p,fo}}{10^6\;{\rm GeV}^{-1}}\right)^{\frac{1}{2}}\left(\frac{B_{\rm p,fo}}{10^{-3}\;{\rm GeV}^2}\right) \simeq 5\times\epsilon_\mathrm{fo}^{-\frac{1}{2}}. }
    % to be more precise, 4.510

\subsection{Baryon isocurvature perturbation}
\label{sec:Formulas}
Now we are ready to express the baryon isocurvature perturbations in the wavenumber space,
in terms of the magnetic field power spectrum.
An expression in the position space is often useful as well. We explicitly derive it in Appendix  \ref{sec:realspacecalG}. 

Equation \eqref{eq:ProducedBAU} implies that the spatial distribution of the BAU at its freeze out, $\eta_{B,{\rm fo}}$, is proportional to 
${\bm A} \cdot {\bm B}|_{T=T_{\rm fo}}$,
\eq{\label{eq:BAUformula}
    \eta_{B,{\rm fo}}({\bm x})=\mathcal{C}\left.{\bm A}({\bm x}) \cdot {\bm B}({\bm x})\right|_{T=T_{\rm fo}} = \mathcal{C}\left.({\bm A} \cdot {\bm \nabla} \times {\bm A})({\bm x})\right|_{T=T_{\rm fo}},
    }
whose Fourier mode is expressed as
\eq{\label{eq:eta(q)}
    \eta_{B,{\rm fo}}({\bm k})\equiv\int d^3r e^{-i{\bm k}\cdot{\bm r}}\eta_{B,{\rm fo}}({\bm r})
    =i\epsilon_{ijk}\mathcal{C}\int\frac{d^3p}{(2\pi)^3}A_i^*({\bm p}-{\bm k})p_jA_k({\bm p}).
    }
To investigate the distribution of the baryon-number fluctuation, 
$\delta \eta_B({\bm x}) \equiv \eta_B({\bm x}) - {\overline \eta}_B$, 
we use two point function of the relative fluctuation, $S_B(\boldsymbol{x}) \equiv \delta \eta_B({\bm x})/{\overline \eta}_B$, expressed as
\eq{\label{eq:PtbCorr}
    \mathcal{G}({\bm r})
    \equiv\langle S_{B,{\rm fo}}(\boldsymbol{x})S_{B,{\rm fo}}(\boldsymbol{x}+\boldsymbol{r})\rangle
    =\frac{\langle \eta_{B,{\rm fo}}(\boldsymbol{x})\eta_{B,{\rm fo}}(\boldsymbol{x}+\boldsymbol{r})\rangle}{{\overline \eta}_B^2} - 1.
}
Then its Fourier transform or the power spectrum is given by
\begin{align}\label{eq: calG}
    \mathcal{G}({\bm k})
    &=\frac{1}{V}\frac{\langle \left| \eta_{B,{\rm fo}}({\bm k})\right|^2 \rangle}{{\overline \eta}_B^2} - (2\pi)^3\delta^3({\bm k})\notag\\
    &=\epsilon_{ijk}\epsilon_{lmn}\frac{\mathcal{C}^2}{V{\overline \eta}_B^2}\int\frac{d^3p}{(2\pi)^3}\int\frac{d^3p'}{(2\pi)^3}p_jp'_m
    J_{ikln}({\bm k}-{\bm p},{\bm p},{\bm k}-{\bm p}',{\bm p}')- (2\pi)^3\delta^3({\bm k}),
\end{align}
where
\begin{align}\label{eq:Jdef}
    J_{ikln}({\bm q}_1,{\bm q}_2,{\bm q}_3,{\bm q}_4)&\equiv\langle A_i^*({\bm q}_1)A^*_k({\bm q}_2)A_l({\bm q}_3)A_n({\bm q}_4)\rangle\notag\\
    &=\langle A^*_i({\bm q}_1)A^*_k({\bm q}_2)\rangle\langle A_l({\bm q}_3)A_n({\bm q}_4)\rangle
    +\langle A^*_i({\bm q}_1)A_l({\bm q}_3)\rangle\langle A^*_k({\bm q}_2)A_n({\bm q}_4)\rangle\notag\\
    &\hspace{56.3mm}+\langle A^*_i({\bm q}_1)A_n({\bm q}_4)\rangle\langle A^*_k({\bm q}_2)A_l({\bm q}_3)\rangle\notag\\
    &=(2\pi)^6\left[
    \delta^3({\bm q}_1+{\bm q}_2)\delta^3({\bm q}_3+{\bm q}_4)\mathcal{F}^A_{ik}({\bm q}_1)\mathcal{F}^A_{ln}(-{\bm q}_3)\right.\notag\\
    &\hspace{26mm}+\delta^3({\bm q}_1-{\bm q}_3)\delta^3({\bm q}_2-{\bm q}_4)\mathcal{F}^A_{il}({\bm q}_1)\mathcal{F}^A_{kn}({\bm q}_2)\notag\\
    &\hspace{26mm}+\left.\delta^3({\bm q}_1-{\bm q}_4)\delta^3({\bm q}_2-{\bm q}_3)\mathcal{F}^A_{in}({\bm q}_1)\mathcal{F}^A_{kl}({\bm q}_2)\right]. 
\end{align}
Here we have used the reality condition of the gauge fields, $A_i^*({\bm k}) = A_i(-{\bm k})$. 
In the second equality,
we have decomposed the four-point function of the vector potential into pairs of two-point functions, assuming the Gaussian distribution.
The contribution of each term to Eq. (\ref{eq: calG}) is evaluated as follows. 
The first term in the last line of Eq. (\ref{eq:Jdef}) is calculated as
\begin{align}\label{eq:calG1}
    \epsilon_{ijk}\epsilon_{lmn}\frac{(2\pi)^6\mathcal{C}^2}{V{\overline \eta}_B^2}
    \int\frac{d^3p}{(2\pi)^3}\int\frac{d^3p'}{(2\pi)^3}p_jp'_m
    \left[\delta^3({\bm k})\right]^2\mathcal{F}^A_{ik}({\bm k}-{\bm p})\mathcal{F}^A_{ln}({\bm p}'-{\bm k}&)\notag\\
    \quad=\frac{(2\pi)^3\mathcal{C}^2}{{\overline \eta}_B^{2}}\left[\frac{1}{\pi^2}\int dp p^3 A(p)\right]^2\delta^3({\bm k})
    =(2\pi)^3\delta^3({\bm k}&).
\end{align}
This contribution is nothing but the homogeneous component of the baryon asymmetry
and cancels with the last term in Eq. (\ref{eq: calG}).
It is generated only from the antisymmetric part of the power spectrum.
On the other hand, the remaining contributions, which generate the inhomogeneities of the baryon asymmetry,
come from both symmetric and antisymmetric parts.
This can be seen by explicitly calculating 
the second term together with the third term as
\begin{align}\label{eq:calG23}
    \epsilon_{ijk}\epsilon_{lmn}&\frac{(2\pi)^6\mathcal{C}^2}{V{\overline \eta}_B^2}
    \int\frac{d^3p}{(2\pi)^3}\int\frac{d^3p'}{(2\pi)^3}p_jp'_m
    \left\{\left[\delta^3({\bm p}-{\bm p}')\right]^2\mathcal{F}^A_{il}({\bm k}-{\bm p})\mathcal{F}^A_{kn}({\bm p})\right.\notag\\
    &\hspace{50mm}+\left.\left[\delta^3({\bm k}-{\bm p}-{\bm p}')\right]^2\mathcal{F}^A_{in}({\bm k}-{\bm p})\mathcal{F}^A_{kl}({\bm p})\right\}\notag\\
    =&\frac{\mathcal{C}^2}{{\overline \eta}_B^2}
    \int\frac{d^3p}{(2\pi)^3}\left[p^2S(|{\bm k}-{\bm p}|)S(p)+|{\bm k}-{\bm p}|pA(|{\bm k}-{\bm p}|)A(p)\right] \notag \\
    &\hspace{50mm} \times \left[1-\frac{2({\bm k}-{\bm p})\cdot{\bm p}}{p^2}+\frac{\left(({\bm k}-{\bm p})\cdot{\bm p}\right)^2}{|{\bm k}-{\bm p}|^2p^2}\right]. 
\end{align}
Putting Eqs. (\ref{eq:calG1}) and (\ref{eq:calG23}) into Eq. (\ref{eq: calG}),
we obtain
\begin{align}
\label{eq:calG}
    \mathcal{G}({\bm q})
    &=\frac{\mathcal{C}^2}{{\overline \eta}_B^2}
    \int\frac{d^3p}{(2\pi)^3}\left[p^2S(|{\bm k}-{\bm p}|)S(p)+|{\bm k}-{\bm p}|pA(|{\bm k}-{\bm p}|)A(p)\right] \notag \\
    &\hspace{50mm} \times \left[1-\frac{2({\bm k}-{\bm p})\cdot{\bm p}}{p^2}+\frac{\left(({\bm k}-{\bm p})\cdot{\bm p}\right)^2}{|{\bm k}-{\bm p}|^2p^2}\right].
\end{align}
The essential point, which has already been noticed in Ref.~\cite{Giovannini:1997eg}, is that 
the non-helical part $S(k)$ contributes to the two-point correlation function of the magnetic helicity.
This is because $A(k)=0$ does not mean that the helicity,  ${\bm A} \cdot {\bm B}$, vanishes 
at every position for $S(k)\not = 0$. 
Thus the two-point correlation function of the magnetic helicity yields a non-vanishing value even for $A(k)=0$, 
which plays crucial role in constraining the primordial hypermagnetic fields.

\section{Observational constraints}
\label{sec:AfterBG}
The baryon isocurvature perturbation, which we explored in the previous section, 
imposes constraints on the baryogenesis scenario from the hypermagnetic helicity decay, 
in the parameter space where the homogeneous part of the BAU can be explained. 
In this section, we examine these constraints in terms of the magnetic field properties. 
We also briefly summarize the implication of the observation of the intergalactic magnetic fields 
on the scenario, adopting the magnetic field evolution suggested by the MHD.

\subsection{Constraints from inhomogeneous Big Bang Nulceosynthesis}
\label{sec:constraint from BBN}
Let us first examine the constraint from the baryon isocurvature perturbations. 
It is well-known that they are strongly constrained by the CMB~\cite{Akrami:2018odb, Hu:1994tj, Dent:2012ne, Kawasaki:2007mb, Kawasaki:2014fwa}.\footnote{In addition, future 21 cm line observations are expected to yield constraints on the baryon 
isocurvature perturbations on smaller scales~\cite{Kawasaki:2011ze, Takeuchi:2013hza, Sekiguchi:2013lma}.}
However, it is sensitive only to the comoving scales larger than 10 Mpc. 
In the scenario of the baryogenesis from hypermagnetic helicity decay, baryon isocurvature perturbations
are produced on much smaller scales, since they correspond to the magnetic field coherence length as we have seen in the 
previous section, which depends on the magnetogenesis mechanism. 
Thus the constraints from the CMB observations are generally not applicable. 

Fortunately,  baryon isocurvature perturbation on much smaller scales 
can be constrained by the inhomogeneous BBN~\cite{Inomata:2018htm}. 
Relatively small-scale baryon number fluctuations,
larger than the comoving neutron diffusion scale at the BBN, $k_{\mathrm{d}}^{-1}\simeq 0.0025\;\mathrm{pc}$,
should not be very large;
otherwise it changes the predictions of the primordial abundance of light elements and spoils the success of the BBN.
Thus we can rule out too large baryon isocurvature perturbations at a smaller scale than the scales
constrained by the CMB observations. 
In contrast, a large inhomogeneity on  scales smaller than $k_\mathrm{d}^{-1}$ at the BBN can be 
smeared out by the neutron diffusion, and the constraints get weaker. 

In our scenario, the typical scale of the baryon isocurvature perturbation is implemented 
in the magnetic field coherence length, characterized by $\xi_\mathrm{c}$, 
as we have seen in the previous section. 
Since at the scale $\sim \xi_\mathrm{c}$ the amplitude of the baryon isocurvature perturbation peaks to reach ${\cal O}(1)$
at the EWSB, 
scenarios with large coherence length, $\xi_\mathrm{c} \gg k_{\mathrm{d}}^{-1}$,
would create antibaryon domains in the observable Universe at the BBN
and are not allowed.
Such a large  coherence length is realized for the hypermagnetic fields produced in inflationary magnetogenesis
or hypermagnetic fields in the MHD turbulence with relatively large field strength.
On the contrary, for smaller coherence length, $\xi_\mathrm{c} \lesssim k_{\mathrm{d}}^{-1}$,
baryon isocurvature perturbation at the BBN leads to the inhomogeneous BBN (IBBN),
explored in Refs.~\cite{Mat:1990, Kainulainen:1998vh, Lara:2005zt, Lara:2006cd, Matsuura:2005rb}. 
In this case,
neutrons diffuse and damp large baryon isocurvature perturbations at the scales smaller than the neutron diffusion scale,
and there remains tiny but non-vanishing baryon isocurvature perturbations
at the scales larger than the neutron diffusion scale, 
from which one can constrain the magnetic field distributions. 
By treating the baryon inhomogeneities at the BBN that survives the smearing effect of the neutron diffusion as a linear perturbation, 
one can estimate the deuterium abundance from the second order effect of perturbations to obtain a constraint on the baryon isocurvature perturbations
to avoid the deutrium overproduction.
The condition for the volume average of the baryon isocurvature perturbation is found to be~\cite{Inomata:2018htm}
\eq{\label{eq:BBNConstraint2}
    \overline{S^{2}_{B,\mathrm{BBN}}({\bm x})}
    <0.016\quad(2\sigma),
    }
where $S_{B,\mathrm{BBN}}({\bm x})$ is the smoothed baryon isocurvature perturbation at the BBN. 

To take into account the smearing effect of the neutron diffusion, 
we 
introduce a Gaussian window function so that the baryon perturbation $\delta\eta_{B}$ at the BBN is evaluated as
\eq{ \label{eq:pertwindow}
    \delta\eta_{B, \mathrm{BBN}}(\boldsymbol{r})
    =\int d^3x W_{D}(\boldsymbol{x}-\boldsymbol{r}) \delta\eta_{B,{\rm fo}}(\boldsymbol{x}),
    \quad W_{D}(\boldsymbol{r}) \equiv \left(\frac{D}{\pi}\right)^{\frac{3}{2}}e^{-D\boldsymbol{r}^2},\quad D\equiv\frac{3}{2}k^2_{\rm d}. 
}
The choice of the window function is motivated by the fact that the neutron diffusion equation is
approximated by a heat equation~\cite{Applegate:1987hm} whose (three-dimensional) heat kernel is the Gaussian function.
Thus we employ $\eta_{B,{\rm fo}}(\boldsymbol{r})$ as the initial condition and just convolute it with the heat kernel $W(\boldsymbol{r})$ to obtain $\eta_{B,\mathrm{BBN}}(\boldsymbol{r})$. The heat kernel is nothing but the window function, seen in the wavenumber space 
and is normalized as $\int d^3 x W_{D}(\boldsymbol{x})=1$, or equivalently $W_{D}(\boldsymbol{k}=0)=1$ in the wavenumber space. We set the width of the Gaussian function so that the root mean square radius $\left(\int d^3 x W_{D}(\boldsymbol{x}) x^2\right)^{1/2}=\sqrt{3/(2D)}$ coincides with $k_{\mathrm{d}}^{-1}$, obeying the definition of the word ``diffusion length'' in Ref. \cite{Applegate:1987hm}. 

From Eq.~\eqref{eq:pertwindow}, the volume average of the smoothed baryon isocurvature perturbation is evaluated as 
\begin{align}
    \label{eq:BBNconstraint3}
    \overline{S^{2}_{B,\mathrm{BBN}}}
    &=\frac{\langle\delta\eta^2_{B,{\rm BBN}}({\bm x})\rangle}{\overline{\eta}_B^2}=\int d^3x W_{D/2}({\bm x})\mathcal{G}({\bm x})\\
    &=\int\frac{d^3k}{(2\pi)^3}e^{-\frac{k^2}{2D}}\mathcal{G}({\bm k})
    \left(<0.016\right).
\end{align}

By inserting Eq. (\ref{eq:calG}) into the last line,
we obtain a formula that directly relates the power spectrum of the magnetic field to the baryon isocurvature perturbations at the BBN.
\begin{align}\label{eq:calGformula}
    &\overline{S^{2}_{B,\mathrm{BBN}}}
    =\int\frac{d^3k}{(2\pi)^3}e^{-\frac{k^2}{2D}}\mathcal{G}({\bm k})\notag\\
    &\,=\frac{\mathcal{C}^2}{{\overline \eta}_B^2}
    \int\frac{d^3k}{(2\pi)^3}\int\frac{d^3p}{(2\pi)^3}e^{-\frac{k^2}{2D}}\left[p^2S(|{\bm k}-{\bm p}|)S(p)+|{\bm k}-{\bm p}|pA(|{\bm k}-{\bm p}|)A(p)\right] \notag \\
    &\hspace{50mm} \times \left[1-\frac{2({\bm k}-{\bm p})\cdot{\bm p}}{p^2}+\frac{\left(({\bm k}-{\bm p})\cdot{\bm p}\right)^2}{|{\bm k}-{\bm p}|^2p^2}\right]\notag\\
    &\,=\frac{\mathcal{C}^2}{4\pi^4{\overline \eta}_B^2}
    \int dk_1 dk_2 k_1^2k_2^2\sum_{\pm}\left( \pm \left\{ \frac{(k_1\pm k_2)^2}{2} \left[S(k_1)S(k_2)\pm A(k_1)A(k_2)\right]\frac{D}{k_1k_2}\left(1\mp\frac{D}{k_1k_2} \right)\right.\right.\notag\\
    &\hspace{15mm}
    +\left. \left[\frac{k_1^2+k_2^2}{2}S(k_1)S(k_2)+k_1k_2 A(k_1)A(k_2)\right]\left(\frac{D}{k_1k_2}\right)^3 \right\}
  \left.\exp \left[ -\frac{(k_1\mp k_2)^2}{2D}\right]\right).
\end{align}
For a given magnetic field spectra, $A(k)$ and $S(k)$, with Eqs.~\eqref{eq:BBNConstraint2} and \eqref{eq:calGformula}
one can determine the constraints on their parameters, which is the main result of the present paper. 
Here we do not assume that the homogeneous part of the baryon asymmetry, $\overline{\eta}_B$, 
is produced by the hypermagnetic helicity decay. Therefore, we can adopt Eq.~\eqref{eq:calGformula} 
also for the case where other baryogenesis mechanisms are responsible for the net BAU, which will be investigated in \S~\ref{sec:Application}. 

We also note that there is another issue at the BBN on the constraint of the magnetic field properties. 
Mainly because magnetic fields contribute to the energy density of the Universe 
as an additional relativistic degree of freedom at the time of the BBN, 
which are constrained in terms of the effective numbers of neutrino species, 
the magnetic field energy density or the field strength is constrained
as $B_{\mathrm{c,BBN}}\lesssim10^{-6}\;\mathrm{G}$~\cite{Grasso:2000wj}
irrespective to its coherent length.

\subsection{Magnetic field evolution until today and implications from the intergalactic magnetic fields}
\label{sec:Magnetic field evolution}
The long-range (hyper)magnetic fields 
generated in the radiation dominated era (including before the EWSB) will evolve according to the MHD equations
and can remain until today as the IGMFs. 
Thus, the constraints obtained in the way we developed in the previous subsection 
can be rewritten in terms of the parameters of the IGMFs. 
In this subsection, we give the formulae that connect the properties of the magnetic fields at 
the EWSB and the present.

With sufficiently strong magnetic fields, the bulk velocity fields are excited due to the MHD, 
and the system enters the turbulence regime once the eddy turnover scale of the velocity fields catches up 
the coherence length of the magnetic fields. 
Before that the magnetic fields evolve adiabatically (the spectrum is unchanged in terms of the comoving quantities) 
whereas they evolve according to the scaling law of the direct or inverse cascade process
until the recombination~\cite{Banerjee:2004df}.
We assume that the magnetic fields evolve adiabatically again after the recombination. 

More concretely, we treat the magnetic field evolution as follows. 
At the EWSB the coherence length of the magnetic fields should be longer than the eddy turnover scale \cite{Banerjee:2004df, Durrer:2013pga},
\begin{equation} \label{eq:catchup}
     \xi_\mathrm{p}|_{T=T_\mathrm{fo}} >  \xi_\mathrm{p,ed}|_{T=T_\mathrm{fo}} \simeq \left.v_A\right|_{T=T_\mathrm{fo}} t_{\rm fo} = \left.\left(\frac{2 {\cal E}_\mathrm{p}}{\rho_\mathrm{p}+p_\mathrm{p}}\right)^{\frac{1}{2}} \right|_{T=T_\mathrm{fo}}t_{\rm fo},  
\end{equation}
where $\rho_\mathrm{p}$ and $p_\mathrm{p}$ are the physical energy density and pressure of the charged plasma particles, respectively, 
and the subscript p represents that the quantity is the physical one. 
Here we have assumed that the fluid velocity reaches at the Alfven velocity $v_A = \sqrt{2 {\cal E}_\mathrm{p}/(\rho_\mathrm{p}+p_\mathrm{p})} =  \sqrt{2 {\cal E}_\mathrm{c}/(\rho_\mathrm{c}+p_\mathrm{c})}$, 
which indicates the equipartition between the magnetic fields and fluid velocity fields.  
Note that it is convenient to investigate with physical quantities.
If $\xi_\mathrm{p}|_{T=T_\mathrm{fo}} = \xi_\mathrm{p,ed}|_{T=T_\mathrm{fo}}$, this suggests that the magnetic fields have already started to 
evolve according to the cascade process at the EWSB.
If the coherence length is longer than the eddy turnover scale at the EWSB, 
eventually the eddy turnover scale catches up the coherence length of the magnetic fields at
\begin{equation}
    t = t_\mathrm{sc} \equiv \left.\frac{\xi_\mathrm{p}}{\sqrt{2 {\cal E}_\mathrm{p}/(\rho_\mathrm{p}+p_\mathrm{p})}}\right|_{t=t_\mathrm{sc}}, 
\end{equation}
and the cascade process is started. Note that if $t_\mathrm{rec}<t_\mathrm{sc}$, with $t_{\rm rec}$ being the physical cosmic time at the recombunation,
the system never enters the cascade regime and evolve always adiabatically in the entire cosmic history after magnetogenesis.

The scaling law of the cascade process of the magnetic field evolution depends on the complicated 
properties of the distributions of the magnetic fields as well as the fluid velocities, 
which is often hard to determine. 
Instead of identifying the scaling law with such properties, with turning back to the comoving quantities, 
let us parameterize the scaling evolution of the magnetic fields, following Ref.~\cite{Brandenburg:2016odr}, as 
\begin{equation}
    {\cal E}_\mathrm{c} \propto \tau^{-\frac{2(\beta+1)}{\beta+3}}, \quad \xi_\mathrm{c} \propto  \tau^{\frac{2}{\beta+3}},
\end{equation} 
where $\tau$ is the conformal time,
and $\beta$ is a parameter determined by the initial conditions of the magnetic fields and the fluid velocity fields.
Note however that the magnetic field spectrum alone does not fix the value of $\beta$.~\footnote{$\beta$ is the parameter that quantifies the decay of the spectral energy around the comoving wave number $k = \xi_c^{-1}$~\cite{Brandenburg:2016odr}.
It is an independent parameter of $\alpha$ in \S \ref{sec:Power law}, which parameterizes the shape of the magnetic field power spectrum.} 
Such a scaling law has also been considered in Refs.~\cite{Olesen:1996ts, Brandenburg:2017neh}.
The eddy turnover scale evolves with $\xi_\mathrm{p,ed} = t \sqrt{2 {\cal E}_\mathrm{c}(\tau)/(\rho_\mathrm{c}+p_\mathrm{c})} \propto \tau^{2/(\beta+3)}$ in the radiation dominated Universe and hence the relationship $\xi \simeq \xi_\mathrm{ed}$ is continuously satisfied during the cascade regime.
If the magnetic fields are maximally helical, the helicity conservation uniquely determines the scaling law of the magnetic field evolution as $\beta=0$, which corresponds to the so-called ``inverse cascade'' regime.
On the other hand, if the magnetic fields are not maximally helical, there are not apparent conserved quantities
to determine $\beta$,
and hence we here take $\beta (\geq 0)$ as the parameter of the system but do not explore its physical origin. 
In particular, we take $\beta=2$ as a representative value for the non-maximally helical case, 
which is often realized in the numerical simulations in Ref.~\cite{Brandenburg:2017neh}. 
We keep in mind that a smaller $\beta$ is hardly realized.
For partially helical magnetic fields with the helicity fraction less than unity at the EWSB, $|\epsilon_\mathrm{fo}|<1$, 
the system becomes more and more helical with the help of the helicity conservation, and the helicity fraction $\epsilon$ grows 
through the cascade process~\cite{Candelaresi:2011pg}, $\epsilon \propto \tau^{2 \beta/(\beta+3)}$.
If $\epsilon_\mathrm{fo}> (\tau_\mathrm{rec}/\tau_\mathrm{sc})^{-2\beta/(\beta+3)}$, 
the system becomes maximally helical, $\epsilon=1$, before the recombination at $\tau = \tau_\mathrm{sc} \epsilon_\mathrm{fo}^{-(\beta+3)/2 \beta}$, 
and $\beta$ turns to 0. 
If $\epsilon_\mathrm{fo} < (\tau_\mathrm{rec}/\tau_\mathrm{sc})^{-2\beta/(\beta+3)}$, 
the cascade process is described by the single $\beta$ throughout until the recombination
as long as the helicity fraction is less than 1. 
Assuming that the magnetic field evolve adiabatically after the recombination until the present, 
we can express the present magnetic field energy density ${\cal E}_0$ and coherence length $\xi_0$ as
\begin{align} \label{eq:MF1}
    {\cal E}_0 &=\left\{\begin{array}{ll}
    \epsilon_\mathrm{fo}^{\frac{2}{3}}\left(\dfrac{\tau_{\mathrm{sc}}}{\tau_{\mathrm{rec}}}\right)^{\frac{2}{3}}\mathcal{E}_\mathrm{c}|_{T=T_\mathrm{fo}}
    \simeq1.1\times10^{-8}\epsilon_{\rm fo}^{\frac{2}{3}}
    \left(\dfrac{\tau_{\mathrm{sc}}}{\tau_{\mathrm{fo}}}\right)^{\frac{2}{3}}\mathcal{E}_\mathrm{c}|_{T=T_\mathrm{fo}} 
    &\quad \text{for}\quad \epsilon_\mathrm{fo}> \left(\dfrac{\tau_\mathrm{rec}}{\tau_\mathrm{sc}}\right)^{-\frac{2\beta}{\beta+3}}, \vspace{3mm}\\
    \left(\dfrac{\tau_{\mathrm{sc}}}{\tau_{\mathrm{rec}}}\right)^{\frac{2(\beta+1)}{\beta+3}}\mathcal{E}_\mathrm{c}|_{T=T_\mathrm{fo}}
    & \quad \text{for}\quad \epsilon_\mathrm{fo}< \left(\dfrac{\tau_\mathrm{rec}}{\tau_\mathrm{sc}}\right)^{-\frac{2\beta}{\beta+3}},
\end{array}\right.  \\ \label{eq:MF2}
    \xi_0 &= \left\{\begin{array}{ll} 
    \epsilon_\mathrm{fo}^{\frac{1}{3}}\left(\dfrac{\tau_{\mathrm{rec}}}{\tau_{\mathrm{sc}}}\right)^{\frac{2}{3}}\xi_\mathrm{c}|_{T=T_\mathrm{fo}}
    \simeq9.3\times 10^{7}\epsilon_\mathrm{fo}^{\frac{1}{3}} \left(\dfrac{\tau_{\mathrm{fo}}}{\tau_{\mathrm{sc}}}\right)^{\frac{2}{3}} \xi_\mathrm{c}|_{T=T_\mathrm{fo}} 
    &\quad \text{for}\quad \epsilon_\mathrm{fo}> \left(\dfrac{\tau_\mathrm{rec}}{\tau_\mathrm{sc}}\right)^{-\frac{2\beta}{\beta+3}}, \vspace{3mm}\\
    \left(\dfrac{\tau_{\mathrm{rec}}}{\tau_{\mathrm{sc}}}\right)^{\frac{2}{\beta+3}}\xi_\mathrm{c}|_{T=T_\mathrm{fo}}  
    & \quad \text{for}\quad \epsilon_\mathrm{fo}< \left(\dfrac{\tau_\mathrm{rec}}{\tau_\mathrm{sc}}\right)^{-\frac{2\beta}{\beta+3}},
\end{array}\right. \quad % to be more precise, 1.08 & 9.29
\end{align}
where we have used $\tau_{\mathrm{rec}}/\tau_{\mathrm{fo}}=(\tau_{\mathrm{rec}}/\tau_{\mathrm{eq}})\cdot(\tau_{\mathrm{eq}}/\tau_{\mathrm{fo}})\simeq 9.0\times10^{11}$.  % to be more precise, 8.95
The present helicity fraction is then given by
\begin{equation}\label{eq:helicityratiotoday}
    \epsilon_0 = \left\{ \begin{array}{ll} 
    1 &  \quad \text{for}\quad \epsilon_\mathrm{fo}> \left(\dfrac{\tau_\mathrm{rec}}{\tau_\mathrm{sc}}\right)^{-\frac{2\beta}{\beta+3}}, \vspace{3mm}\\ 
    \left(\dfrac{\tau_\mathrm{rec}}{\tau_\mathrm{sc}}\right)^{\frac{2\beta}{\beta+3}} \epsilon_\mathrm{fo} &  \quad \text{for}\quad \epsilon_\mathrm{fo} <  \left(\dfrac{\tau_\mathrm{rec}}{\tau_\mathrm{sc}}\right)^{-\frac{2\beta}{\beta+3}} \quad \text{and} \quad \tau_\mathrm{rec} > \tau_\mathrm{sc}, \vspace{3mm}\\ 
    \epsilon_\mathrm{fo}&  \quad \text{for}\quad \tau_\mathrm{rec} < \tau_\mathrm{sc}. 
    \end{array}\right.
\end{equation}
We shall also use $B_0\equiv \sqrt{2 {\cal E}_0}$ as the typical magnetic field strength today. 

Now we can present predictions on the magnetic field properties today and 
compare them with the baryogenesis from the hypermagnetic helicity decay. 
From Eqs.~\eqref{eq:ProducedBAU_xiER}, \eqref{eq:catchup}, \eqref{eq:MF1}, \eqref{eq:MF2}, and \eqref{eq:helicityratiotoday}, 
we find that the properties of the present magnetic field, $\xi_0$, $B_0$, and $\epsilon_0$
determine the baryon asymmetry produced by the hypermagnetic helicity decay regardless of its evolution history as
\begin{equation} \label{eq:BAU_presentMF}
    \overline{\eta}_{B} \simeq 10^{-10} \epsilon_0 \left(\frac{{\cal C}}{10^{33} \;\mathrm{Mpc}^{-1} \mathrm{G}^{-2}}\right) \left(\frac{\xi_0}{10^{-9} \;\mathrm{Mpc}}\right) \left(\frac{B_0}{10^{-17} \;\mathrm{G}}\right)^2.
\end{equation}
The fiducial value, $\mathcal{C}\sim10^{33}\;\mathrm{Mpc}^{-1} {\rm G}^{-2}$, is taken from Eq.~\eqref{eq:defcalC}. 
The constraints from the baryon isocurvature perturbation in terms of the present magnetic field properties
can be derived in a similar way, but we need to specify the shape of the magnetic field spectrum 
to evaluate concretely. We perform the investigation in the next section. 

Before closing this section, let us summarize the constraints on the cosmological magnetic fields 
other than those from the inhomogeneous BBN.  
First, once the system enters the cascade regime, the relation between the present magnetic field strength and coherence length is 
determined by their values at the recombination as
\begin{equation} \label{eq:MHD}
    \xi_0 = \xi_\mathrm{c,ed, rec} = \left.a^{-1}\left(\frac{2 {\cal E}_\mathrm{c}}{\rho_c+p_c}\right)^{\frac{1}{2}} \right|_{T=T_\mathrm{rec}} t_{\rm rec}
    \simeq 0.6\;\mathrm{pc} \left(\frac{B_0}{10^{-14}\mathrm{G}}\right). % to be more precise, 0.582 pc
\end{equation}
and hence the primordial magnetic fields generated before the recombination 
should have the property, $\xi_0 \gtrsim 0.6 \;\mathrm{pc} (B_0/10^{-14}\;\mathrm{G})$. 
As we have mentioned, since the magnetic fields contribute to additional relativistic energy density of the Universe, 
the BBN gives the upper bound on the comoving magnetic field strength when it occurs, $B_\mathrm{c,BBN}<10^{-6}\;\mathrm{G}$~\cite{Grasso:2000wj}, regardless of the coherence length. 
Moreover, the CMB anisotropy imposes another upper bound on the cosmological magnetic field strength
whose comoving coherence length is longer than Mpc scales as $B_0<10^{-9}\; \mathrm{G}$~\cite{Ade:2015cva,Barrow:1997mj,Blasi:1999hu,Jedamzik:1999bm,Durrer:1999bk,Yamazaki:2012pg,Trivedi:2010gi,Paoletti:2010rx,Shaw:2010ea,Shaw:2009nf} (summarized in Ref.~\cite{Durrer:2013pga}), 
since they would give additional contributions to the CMB anisotropy.

On the other hand, recent gamma-ray observations of blazars suggest the existence of the intergalactic magnetic fields~\cite{Neronov:1900zz, Tavecchio:2010mk, Ando:2010rb, Dolag:2010ni, Essey:2010nd, Taylor:2011bn,Takahashi:2013lba, Finke:2015ona, Biteau:2018tmv}. 
For example, the Fermi collaboration reported the deficit of the secondary GeV cascade photons that should accompany the 
TeV photons from the blazars, which implies the existence of the intergalactic magnetic fields. 
The latest analysis suggests that the observations can be explained if the magnetic fields are strong enough~\cite{Biteau:2018tmv}, 
\begin{equation}\label{eq:blazarconstraint}
    B_0> \left\{\begin{array}{ll}
    3 \times 10^{-16} \;\mathrm{G} & \quad \text{for} \quad \xi_0 > 10 \;\mathrm{kpc}  \\
    3 \times 10^{-16} \;\mathrm{G}  \times \left(\dfrac{\xi_0}{10 \;\mathrm{kpc}}\right)^{-\frac{1}{2}} & \quad \text{for} \quad \xi_0 < 10 \;\mathrm{kpc} 
    \end{array}\right. . 
\end{equation}
Note that Ref.~\cite{Biteau:2018tmv} gives the constraint only for the coherence length $10^2 \;\mathrm{pc}<\xi_0<10^2 \;\mathrm{Mpc}$, 
and hence strictly speaking we can use the constraint only in these ranges, 
but we will use the extrapolated values as the reference. 
Comparing Eq.~\eqref{eq:blazarconstraint} to Eq.~\eqref{eq:BAU_presentMF}, 
the blazar observation can be explained by the primordial magnetic fields responsible for the present BAU
only for $\epsilon_0\lesssim10^{-9}$. 
In particular, if the magnetic fields have entered in the cascade regime before the recombination, 
from Eqs.~\eqref{eq:BAU_presentMF} and \eqref{eq:MHD} we obtain the condition for the BAU to be explained
by the hypermagnetic helicity decay in terms of the present magnetic field properties as
\eq{\label{eq:BAUonETO}
    B_0 \simeq 1\times10^{-17}\;{\rm G}\times\epsilon_0^{-\frac{1}{3}}, \quad 
    \xi_0 \simeq  6 \times 10^{-10}\; \mathrm{Mpc}\times\epsilon_0^{-\frac{1}{3}}, 
    % to be more precise, 1.120, 0.6722
}
regardless of the shape of the spectrum.

\section{Specific models}
\label{sec:Concrete models}
In the previous sections, we have presented a formalism to explain the BAU from the hypermagnetic helicity decay in connection with the properties of the magnetic field both at the EWSB and the present.
It turned out that the scenario is consistent with the IGMFs suggested by the blazar observations only when  
the helicity fraction is very small, $\epsilon \lesssim10^{-9}$. 
From baryon isocurvature perturbations, we can already constrain the magnetic field coherence length
roughly as  $\xi_\mathrm{c,fo} \lesssim k_{\rm d}^{-1}$ to avoid the deuterium overproduction. 
However, to apply the more concrete constraint described in \S~\ref{sec:constraint from BBN}, 
we need to specify the magnetic field spectra at the EWSB. 
In this section,
we specify specific forms of the power spectra, which enable us to extract the general feature of the constraint. 
We also present appropriate constraints in terms of the present magnetic field properties.

\subsection{Delta function}
\label{sec:Delta function}
Let us first investigate 
the simplest choice of $S(k)$ and $A(k)=\epsilon S(k)$ at the EWSB,  a monochromatic form described by the delta function. 
\eq{\label{eq:Delta}
    S(k)=
    \pi^2\frac{B^2_\mathrm{c,fo}}{k_{\sigma}^4} \delta(k-k_\sigma),\quad
    A(k)=\epsilon_\mathrm{fo}S(k),
    }
so that the characteristic properties of the magnetic fields, as well as the net baryon asymmetry are given by
\begin{equation}\label{eq:Delta1}
    {\cal E}_\mathrm{c,fo} = \frac{1}{2} B_\mathrm{c,fo}^2, \quad \xi_\mathrm{c,fo} = k_\sigma^{-1}, \quad h_\mathrm{c,fo} = \epsilon_\mathrm{fo} \xi_\mathrm{c,fo} B_\mathrm{c,fo}^2, \quad \overline{\eta}_B = \epsilon_\mathrm{fo} {\cal C} \xi_\mathrm{c,fo} B_\mathrm{c,fo}^2 , 
\end{equation}
respectively.
Note that the previous studies on the homogeneous part of the baryon asymmetry from the hypermagnetic helicity decay~\cite{Fujita:2016igl,Kamada:2016cnb} 
implicitly assumes such spectra. 

The baryon isocurvature perturbation at the BBN is evaluated from Eq.\  (\ref{eq:calGformula}) as
\eq{\label{eq:Delta_BBN}
    \overline{S^{2}_{B,\mathrm{BBN}}}&=\frac{1}{4}\left(\frac{1}{\epsilon_\mathrm{fo}^2}+1\right)D \xi_\mathrm{c,fo}^{2}\left[2-2D\xi_\mathrm{c,fo}^{2}+D^2 \xi_\mathrm{c,fo}^{4}\left(1-e^{-\frac{2}{D\xi_\mathrm{c,fo}^{2}}}\right)\right].
}
In the limit of large coherence length,
$D\xi_\mathrm{c,fo}^{2} \gg 1$,
\eq{
    \overline{S^{2}_{B,\mathrm{BBN}}}=\frac{1}{3}\left(\frac{1}{\epsilon_\mathrm{fo}^2}+1\right)+\mathcal{O}\left(\left(D\xi_\mathrm{c,fo}^{2}\right)^{-1}\right).
    }
In this limit,
the effect of neutron diffusion is ineffective and the condition~\eqref{eq:BBNConstraint2} is never satisfied.
In the opposite limit,
$D\xi_\mathrm{c,fo}^{2}\ll 1$, 
corresponding to small-scale magnetic fields,
which is the case we are interested in so as to maintain the condition~\eqref{eq:BBNConstraint2},
\eq{\label{eq:Delta2}
    \overline{S^{2}_{B,\mathrm{BBN}}}
    =\frac{1}{2}\left(\frac{1}{\epsilon_\mathrm{fo}^2}+1\right)D\xi_\mathrm{c,fo}^{2}+\mathcal{O}\left(\left(D\xi_\mathrm{c,fo}^{2}\right)^2\right).
}
A factor $\xi_\mathrm{c,fo}^{2}$ appears in the leading term,
convincing us that the shorter the coherence length of the fields is,
the more suppressed the baryon isocurvature perturbations. 
This is because neutrons are more likely to smear out the inhomogeneity of baryon distribution.
Needless to say, the delta-function type power spectrum in the wavenumber space does not mean a 
vanishing correlation on scales larger than the corresponding scale in the position space,
and hence we acquire a suppressed but finite correlation even after smearing out by the neutron diffusion. 

The condition for sufficiently small baryon isocurvature perturbation (Eq. (\ref{eq:BBNConstraint2})), 
with the condition that the present BAU is explained by the hypermagnetic helicity decay (Eq. \eqref{eq:BAUEWSB}) being satisfied,
gives the conditions in terms of the magnetic field properties at the EWSB as well as the present.

First we express the constraint in terms of the magnetic field properties at the EWSB.  The condition for the MHD erasure of the small-scale magnetic fields (Eq.~\eqref{eq:catchup}) is written in terms of the 
physical magnetic field properties as 
\begin{equation}\label{eq:eddyEWSB}
    B_\mathrm{p,fo} \lesssim 6 \times 10^{-3} \;\mathrm{GeV}^2  \times  \left(\frac{\xi_\mathrm{p,fo} }{10^6 \;\mathrm{GeV}^{-1}}\right),
\end{equation}
which does not depend on the helicity fraction $\epsilon_\mathrm{fo}$. 
Here we have used the number of charged relativistic degrees of freedom $g_{*, \mathrm{fo}}^\mathrm{ch}=82.75$.

The condition for the magnetic field to generate the present BAU, 
Eq.~\eqref{eq:BAUEWSB}, 
yields the lower bound of the magnetic field strength together with the 
requirement $\epsilon_\mathrm{fo} \leq 1$, as
\begin{equation} \label{eq:forBAUatEWSB}
    B_\mathrm{p,fo} \gtrsim 5 \times 10^{-3} \;\mathrm{GeV}^2 \left(\frac{\xi_\mathrm{p,fo}}{10^6 \;\mathrm{GeV}^{-1}}\right)^{-\frac{1}{2}}. 
\end{equation}
Conversely, the condition \eqref{eq:BAUEWSB} allow us to determine the helicity fraction $\epsilon_\mathrm{fo}$ for a given magnetic field strength and coherence length. 
This leads to the lower bound of the magnetic field strength from the baryon isocurvature constraint (Eq.~ \eqref{eq:BBNConstraint2})
with Eq.~\eqref{eq:Delta2} as
\begin{equation}\label{eq:BarIsoatEWSB}
    B_\mathrm{p} \lesssim 3 \times 10^{-3} \;\mathrm{GeV}^2 \left(\frac{\xi_\mathrm{p,fo}}{10^{10} \;\mathrm{GeV}^{-1}}\right)^{-1}, 
\end{equation}
for $\epsilon_\mathrm{fo} \ll 1$, 
where we have used the fact that the neutron diffusion scale at the BBN corresponds to the proper wavenumber  $k_\mathrm{d,fo}^\mathrm{p} =   4.4 \times 10^{-15} \;\mathrm{GeV}$ at the EWSB.

\begin{figure}[htbp]
\begin{center}
    \includegraphics[width=0.48\textwidth]{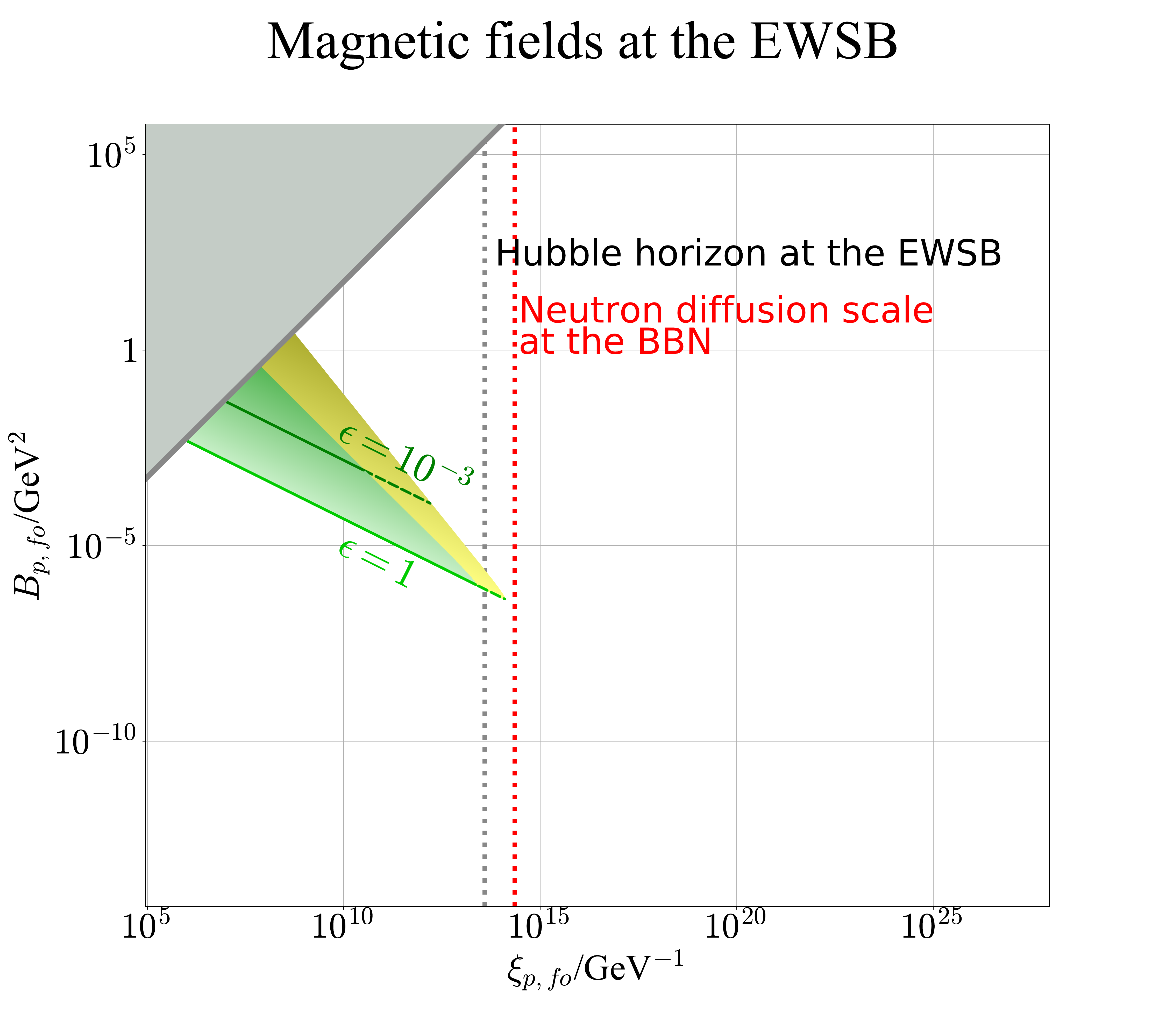}
    \includegraphics[width=0.48\textwidth]{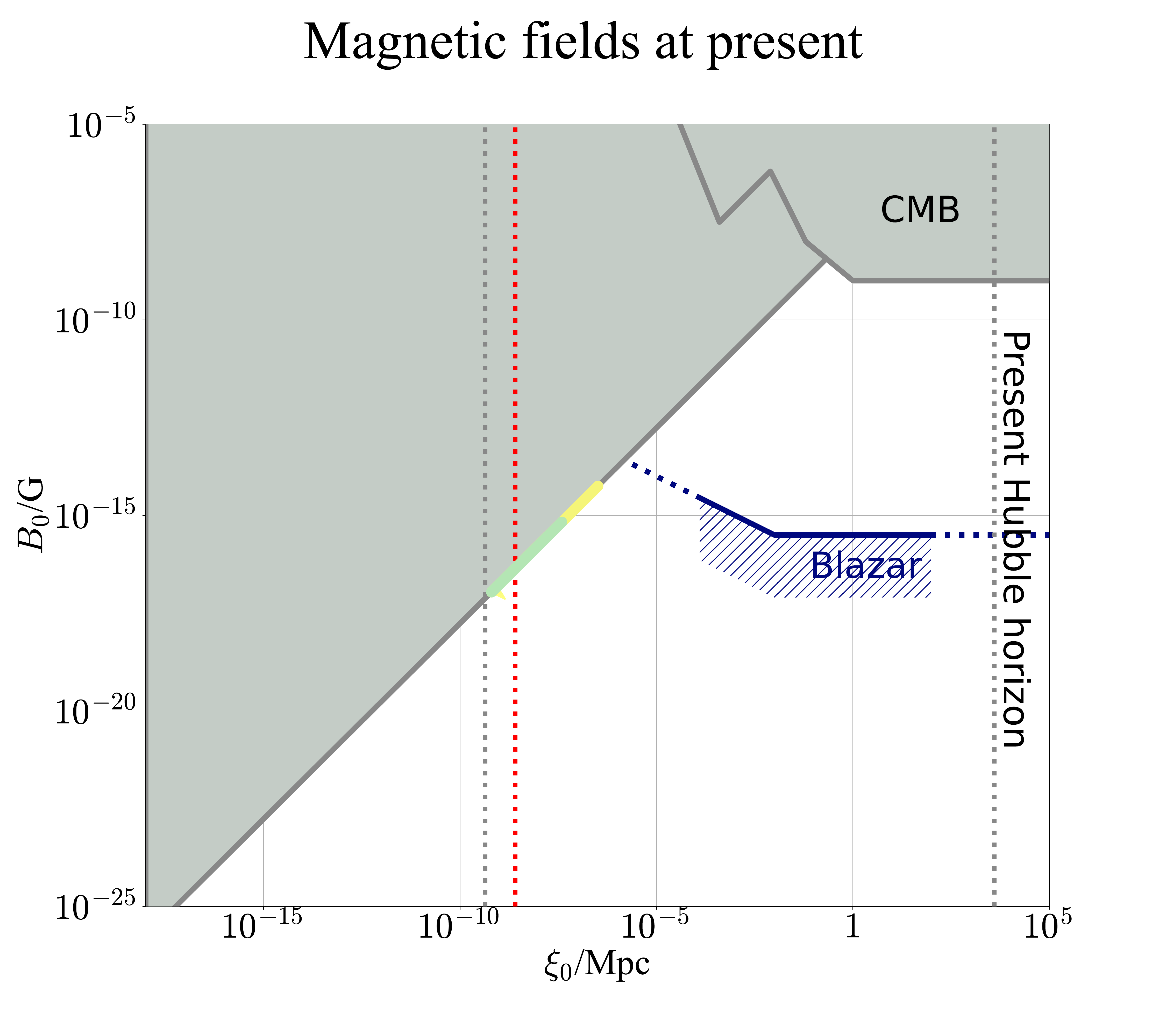}
\caption{\label{fig:1}
    {\it Left.} 
    Constraints on the magnetic fields at the EWSB, Equations.~\eqref{eq:eddyEWSB}, \eqref{eq:forBAUatEWSB},  \eqref{eq:BarIsoatEWSB}, 
    and \eqref{eq:BarIsoatEWSB2} are shown in terms of the physical quantities.
    {\it Right.} 
    Constraints on the present magnetic fields, Eqs. \eqref{eq:BAUonETO}, \eqref{eq:DeltaNow}, and \eqref{eq:PowerNow} are shown. 
    The navy thick line,
    with its lower side hatched, 
    is the lower bounds suggested by blazar observations \cite{Biteau:2018tmv}.
    The navy dotted lines are an extrapolation of Ref. \cite{Biteau:2018tmv}.
    In both panels,
    the gray shaded regions above the eddy turnover scales are the parameter space inconsistent with the MHD evolution.
    The green shaded region and line are the allowed parameter regions for the delta-function model,
    while the yellow shaded region and line are the one for the power-law model with $\alpha=0$.
}
\end{center}
\end{figure}

The parameter space constrained by Eqs.~\eqref{eq:eddyEWSB}, \eqref{eq:forBAUatEWSB}, and \eqref{eq:BarIsoatEWSB} 
(with the behavior at $\epsilon_\mathrm{fo} \sim 1$ being appropriately taken into account)
is shown as the green shaded region of the right panel of Fig.~\ref{fig:1}. 
Here the smallest helicity fraction that satisfies all the conditions is found to be 
$\epsilon_\mathrm{fo} \simeq 2\times 10^{-6}$. 

Next, let us turn to the constraints expressed by  the present IGMF properties. 
First of all, large-scale magnetic fields whose coherence length is larger than the one for the 
MHD turbulence (Eq.~\eqref{eq:MHD}) are ruled out. 
We have seen that such magnetic fields have evolved adiabatically with the comoving coherence length being unchanged. 
Then, it can be easily checked that 
it is impossible to satisfy $\xi_\mathrm{c,fo} < k_\mathrm{d}^{-1}$ , which is required by the baryon isocurvature constraints (Eq.~\eqref{eq:BBNConstraint2}), 
simultaneously satisfying the condition for the generation of the present BAU (Eq.~\eqref{eq:BAU_presentMF}). 
For the IGMF whose properties lie in the line Eq.~\eqref{eq:MHD},  
there remains a possibility to satisfy the baryon isocurvature constraint 
since it has experieced the cascade process and the coherence length is much shorter at the EWSB. 
Since the relationship between the coherence length at the EWSB and that at the present depends on the value of $\beta$, 
the baryon isocurvature constraint today also depends on $\beta$. 
If $\beta>1$, the magnetic fields that satisfy the constraint at the EWSB (green shaded region in Fig.~\ref{fig:1}) 
become maximally helical until the recombination. 
This can be seen that in the $B$-$\xi$ figure, the magnetic field  
evolves with $B_\mathrm{c} \propto \xi_\mathrm{c}^{-(\beta+1)/2}$ whereas the baryon isocurvature constraint 
yields $B_\mathrm{c} \propto \xi_\mathrm{c}^{-1}$ (Eq.~\eqref{eq:BarIsoatEWSB}). 
In this case, the IGMF property is uniquely determined as
\begin{equation} \label{eq:maxhelIGMFBAU}
    B_0 \simeq 1 \times 10^{-17} \;\mathrm{G}, \quad \xi_0 \simeq 6 \times 10^{-4} \;\mathrm{pc}. 
\end{equation}
The longest coherence length is obtained for the extreme choice, $\beta=0$, (which would be hard to realize, though), 
where the cascade is most effective. 
In such a case, the helicity fraction does not change so that  the smallest one is  
$\epsilon_0\sim\epsilon_{\rm fo}\sim 2 \times 10^{-6}$. 
This uniquely determines the upper bound of the magnetic field strength and coherence length for $0\leq \beta \leq 1$ as
\eq{ \label{eq:DeltaNow}
    1\times10^{-17}\;\mathrm{G} \lesssim B_0  \lesssim 7 \times 10^{-16}\;\mathrm{G} \times \left(\dfrac{\tau_{\rm rec}}{\tau_{\rm fo}}\right)^{-\frac{\beta+1}{\beta+3}+\frac{1}{3}},\;
    6\times 10^{-4} \;\mathrm{pc}  \lesssim  \xi_0  \lesssim 0.07 \;\mathrm{pc}\times \left(\dfrac{\tau_{\rm rec}}{\tau_{\rm fo}}\right)^{\frac{2}{\beta+3}-\frac{2}{3}},
    }
with $\tau_\mathrm{rec}/\tau_\mathrm{fo} \simeq 9\times 10^{11}$
where the lower bound is for the case $\epsilon_0=1$. 
The allowed region is depicted as the green line in the right panel of Fig.~\ref{fig:1}.
The constraint \eqref{eq:DeltaNow} falls below the blazar constraint (Eq.~\eqref{eq:blazarconstraint}), $B_0\lesssim3\times10^{-14}$ G on the line of the MHD turbulence (Eq.~\eqref{eq:MHD}), by a factor of $\mathcal{O}(10^2)$.

\subsection{A Power-law spectrum with an exponential cutoff}
\label{sec:Power law}
Next we study a more realistic form of the spectrum, a power-law spectrum with an exponential cutoff, 
\eq{\label{eq:PLmodel}
    S(k)
    =\frac{2\pi^2}{\Gamma(\frac{5\;+\;\alpha}{2})}
    \frac{B_\mathrm{c,fo}^2}{k_{\sigma}^5} 
    \left(\frac{k}{k_\sigma}\right)^{\alpha} \exp \left[- \left(\frac{k}{k_\sigma}\right)^2\right],\quad
    A(k)=\epsilon_\mathrm{fo}S(k).
    }
In this case, characteristic properties of the magnetic fields,
as well as the net baryon asymmetry produced by the fields, are given by
\begin{equation}
    {\cal E}_\mathrm{c,fo} =
    \frac{1}{2} B_\mathrm{c,fo}^2, \quad
    \xi_\mathrm{c,fo} = \left(\zeta_{\alpha} k_\sigma\right)^{-1}, \quad
    h_\mathrm{c,fo} = \epsilon_\mathrm{fo} \xi_\mathrm{c,fo} {\cal E}_\mathrm{c,fo}, \quad
    \overline{\eta}_B = \epsilon_\mathrm{fo} {\cal C} \xi_\mathrm{c,fo} B_\mathrm{c,fo}^2 , 
\end{equation}
respectively,
where $\zeta_{\alpha}\equiv \frac{\Gamma\left(\frac{5\;+\;\alpha}{2}\right)}{\Gamma\left(2+\frac{\alpha}{2}\right)}.$
Here we restrict ourselves as $\alpha>-4$ otherwise the energy density and coherence length 
are subject to infrared divergences.
Such a spectrum is introduced, for example, in Ref. \cite{Giovannini:1997gp,Giovannini:1997eg},
and we expect that this single-power law with exponential decay model captures the essential properties of a broader class of spectrum including the one \cite{Brandenburg:2018ptt} that has a Batchelor spectrum at small $k$, a Kolmogorov spectrum at intermediate $k$, and an exponential cutoff at large $k$  because the baryon isocurvature constraints
are mainly determined by the peak and the infrared tail of the spectrum.
Batchelor spectrum cut off at large wavenumber corresponds to $\alpha=0$ in our notation,
so we are mainly interested in the parameter range around $\alpha\sim0$. 

In this case, the baryon isocurvature perturbation at the BBN is given from Eq. (\ref{eq:calGformula}) as
\begin{align}
    \overline{S^{2}_{B,\mathrm{BBN}}}=\frac{(\zeta_\alpha \xi_\mathrm{c,fo})^2D}{2\Gamma\left(2+\frac{\alpha}{2}\right)^2}  \int dq_1&dq_2q_1^{\alpha}q_2^{\alpha}e^{-q_1^2-q_2^2}\notag\\
    \sum_{\pm}&\left(e^{-\frac{(q_1\pm q_2)^2}{2(\zeta_\alpha \xi_\mathrm{c,fo})^2D}}
    \left\{\left(1\mp\frac{1}{\epsilon_\mathrm{fo}^2}\right)(q_1\mp q_2)^2\left[q_1q_2
    \pm (\zeta_\alpha \xi_\mathrm{c,fo})^2D\right] \right.\right.\notag \\
    &\;\; \left.\left.\mp\left(2+\frac{1}{\epsilon_\mathrm{fo}^2}\frac{q_1^2+q_2^2}{q_1q_2}\right)\left[(\zeta_\alpha \xi_\mathrm{c,fo})^2D\right]^2\right\}\right), 
\end{align}
where we have introduced a dimensionless wavenumber $q_i = \zeta_\alpha \xi_\mathrm{c,fo} k_i$. 
In the limit of large coherence length, for $\alpha > -3$, 
$(\zeta_\alpha \xi_\mathrm{c,fo})^2 D\gg 1$,
\eq{\label{eq:PL_BBN_longlimit}
    \overline{S^{2}_{B,\mathrm{BBN}}}
    &=\frac{1}{3}\left[\frac{\Gamma\left(\frac{3\;+\;\alpha}{2}\right)\Gamma\left(\frac{5\;+\;\alpha}{2}\right)}{\Gamma\left(2+\frac{\alpha}{2}\right)^2\epsilon_\mathrm{fo}^2}+1\right]+\mathcal{O}\left(\left[(\zeta_\alpha \xi_\mathrm{c,fo})^2 D\right]^{-1}\right),
}
where the terms with positive powers of $(\zeta_\alpha \xi_\mathrm{c,fo})^2 D$ are canceled out by the exponential factor. 
In the case $\alpha\leq-3$, since the integrand behaves as $k_1^{\alpha+2} k_2^{\alpha+2} \times {\cal O}(k_1^2 , k_2^2)/(\epsilon_\mathrm{fo}^2 (\zeta_\alpha \xi_\mathrm{c,fo})^2 D)$, 
the volume average of the baryon isocurvature perturbation diverges at the infrared as
\begin{equation} \label{eq:IRdiverge}
    \overline{S^{2}_{B,\mathrm{BBN}}} \sim \frac{1}{\epsilon_\mathrm{fo}^2} (\zeta_\alpha \xi_\mathrm{c,fo}
    \Lambda_\mathrm{IR})^{3+\alpha}, 
\end{equation}
where $\Lambda_\mathrm{IR}$ is the infrared cutoff, 
and hence it becomes unacceptably large for $(\zeta_\alpha \xi_\mathrm{c,fo})^2 D\gg1$. 
In this limit, once more, 
the effect of neutron diffusion is ineffective and the condition (\ref{eq:BBNConstraint2}) is never satisfied.

In the other limit $(\zeta_\alpha \xi_\mathrm{c,fo})^2 D \ll 1$, 
which is the case when the condition (\ref{eq:BBNConstraint2}) can be satisfied, 
by using the saddle point approximation, we can approximate 
the volume average of the baryon isocurvature perturbation for $\alpha>-5/2$  as
\eq{\label{eq:Power2}
    \overline{S^{2}_{B,\mathrm{BBN}}}
    =\frac{\sqrt{\pi}2^{-2-\alpha}\Gamma\left(\frac{5}{2}+\alpha\right)}{\Gamma\left(2+\frac{\alpha}{2}\right)^2}\left(\frac{1}{\epsilon_\mathrm{fo}^2}+1\right)\left[(\zeta_\alpha \xi_\mathrm{c,fo})^2 D\right]^{\frac{3}{2}}+\mathcal{O}\left(\left[(\zeta_\alpha \xi_\mathrm{c,fo})^2 D\right]^{\frac{5}{2}}\right). }
A factor $\xi_\mathrm{c,fo}^{-3}$ appears in the leading term,
meaning that the neutron dissipation effect is more effective than that in delta-function model,
in which case the corresponding factor is $k_{\sigma}^{-2}$.
See Appx.~\ref{sec:realspacecalG} for more discussion on the interpretation how these difference appears.

The apparent divergence for $-3<\alpha\leq -5/2$ is due to the breakdown of the validity of the saddle point approximation. 
Since the IR behavior of the integrand is the same as that seen in the case with $(\zeta_\alpha \xi_\mathrm{c,fo})^2 D \gg 1$, 
with $(\zeta_\alpha \xi_\mathrm{c,fo})^{-2} D^{-1}$ being taken as the ``UV'' cutoff,  we can estimate as
\begin{equation}
\overline{S^{2}_{B,\mathrm{BBN}}} \sim \frac{1}{\epsilon_\mathrm{fo}^2} \left[(\zeta_\alpha \xi_\mathrm{c,fo})^2 D\right]^{\frac{3+\alpha}{2}}. 
\end{equation}
The divergence at $\alpha<-3$ is the physical infrared divergence and the volume average of the baryon isocurvature perturbation 
behaves as Eq.~\eqref {eq:IRdiverge} in the same way to the case with $(\zeta_\alpha \xi_\mathrm{c,fo})^2 D \gg1$. 
Hereafter we will focus on the case with $\alpha > -5/2$ for simplicity .

Now we are ready to present the constraints on the properties of the magnetic field both at the EWSB as well as those at the present.
As for the magnetic fields at the EWSB, the constraints derived from the MHD, \eqref{eq:eddyEWSB}, and from baryon-number generation, \eqref{eq:forBAUatEWSB}, apply in the same way as in the  the delta-function spectrum.
Hence the baryon isocurvature constraint can be derived in a similar way to the previous subsection. 
By combining Eqs.~\eqref{eq:BAUEWSB} and \eqref{eq:Power2}, the baryon isocurvature constraint Eq.~\eqref{eq:BBNconstraint3}
is rewritten in a form of the constraint on the magnetic field strength as
\begin{equation} \label{eq:BarIsoatEWSB2}
    B_\mathrm{p,fo} \lesssim 3 \times 10^{-2} \;\mathrm{GeV}^2 \times 2^{\frac{\alpha}{4}} \times \frac{\Gamma\left(2+\frac{\alpha}{2}\right)^{\frac{5}{4}}}{\Gamma\left(\frac{5}{2}+\alpha\right)^{\frac{1}{4}}\Gamma\left(\frac{5\;+\;\alpha}{2}\right)^{\frac{3}{4}}} \times \left(\frac{\xi_\mathrm{p,fo}}{10^{10} \;\mathrm{GeV}^{-1}}\right)^{-\frac{5}{4}}, 
\end{equation}
for $\epsilon_\mathrm{fo} \ll 1$. 
The parameter space for $\alpha=0$ constrained by Eqs.~\eqref{eq:eddyEWSB},  \eqref{eq:forBAUatEWSB}, and \eqref{eq:BarIsoatEWSB2} 
(with the behavior at $\epsilon_\mathrm{fo} \sim 1$ being appropriately taken into account)  is shown in the yellow shaded region 
of the right panel of Fig.~\ref{fig:1}. 
The smallest helicity fraction that satisfies all the conditions is evaluated as 
\begin{equation} \label{eq:minfraction}
    \epsilon_\mathrm{fo} = \epsilon_\mathrm{fo}^\mathrm{min}\equiv  1 \times 10^{-8}  \times 2^{-\frac{\alpha}{3}} \frac{\Gamma\left(\frac{5}{2}+\alpha\right)^{\frac{1}{3}} \Gamma\left(\frac{5\;+\;\alpha}{2}\right)}{ \Gamma\left(2+\frac{\alpha}{2}\right)^{\frac{5}{3}}}. 
\end{equation} 

The constraints on the present magnetic field are also given in the same way as in the  delta-function spectrum. 
We see that most part of the parameter region in which present magnetic fields have never entered the MHD turbulence 
does not satisfy the baryon isocurvature constraints 
and almost only those which lie in the line Eq.~\eqref{eq:MHD} 
have the possibility to satisfy them. 
From the $B$ - $\xi$ figure (Fig.~\ref{fig:1}), we can see that 
the magnetic fields that satisfy the baryon isocurvature constraint at the EWSB becomes maximally helical 
if they evolve  more rapidly than $B_\mathrm{c} \propto \xi_\mathrm{c}^{-5/4}$ (See also Eq.~\eqref{eq:BarIsoatEWSB2}). 
This corresponds to the cascade parameter $\beta > 3/2$.
In such a case, again, the IGMF property is uniquely determined as Eq.~\eqref{eq:maxhelIGMFBAU}. 
The strongest IGMF is obtained for $\beta=0$ and $\epsilon_\mathrm{fo} \sim \epsilon_0 = \epsilon_\mathrm{fo}^\mathrm{min}$, 
which gives the parameter spaces of the IGMF properties that can explain the BAU satisfying the baryon isocurvature constraint 
as
\begin{equation}\label{eq:PowerNow}
    1 \times 10^{-17}\;\mathrm{G} \lesssim B_0 \lesssim 3 \times 10^{-15}\;\mathrm{G}  \times \left(\dfrac{\tau_{\rm rec}}{\tau_{\rm fo}}\right)^{-\frac{\beta+1}{\beta+3}+\frac{1}{3}}, \quad 6 \times 10^{-4} \;\mathrm{pc} \lesssim \xi_0 \lesssim 0.3 \;\mathrm{pc}  \times \left(\dfrac{\tau_{\rm rec}}{\tau_{\rm fo}}\right)^{\frac{2}{\beta+3}-\frac{2}{3}},
\end{equation}
with $\alpha=0$. Note that the constraints are quite insensitive to the value of $\alpha$. 
These parameter space is shown in the yellow line in the right panel of Fig.~\ref{fig:1}. 

\begin{figure}
\begin{center}
      \includegraphics[width=0.45\textwidth]{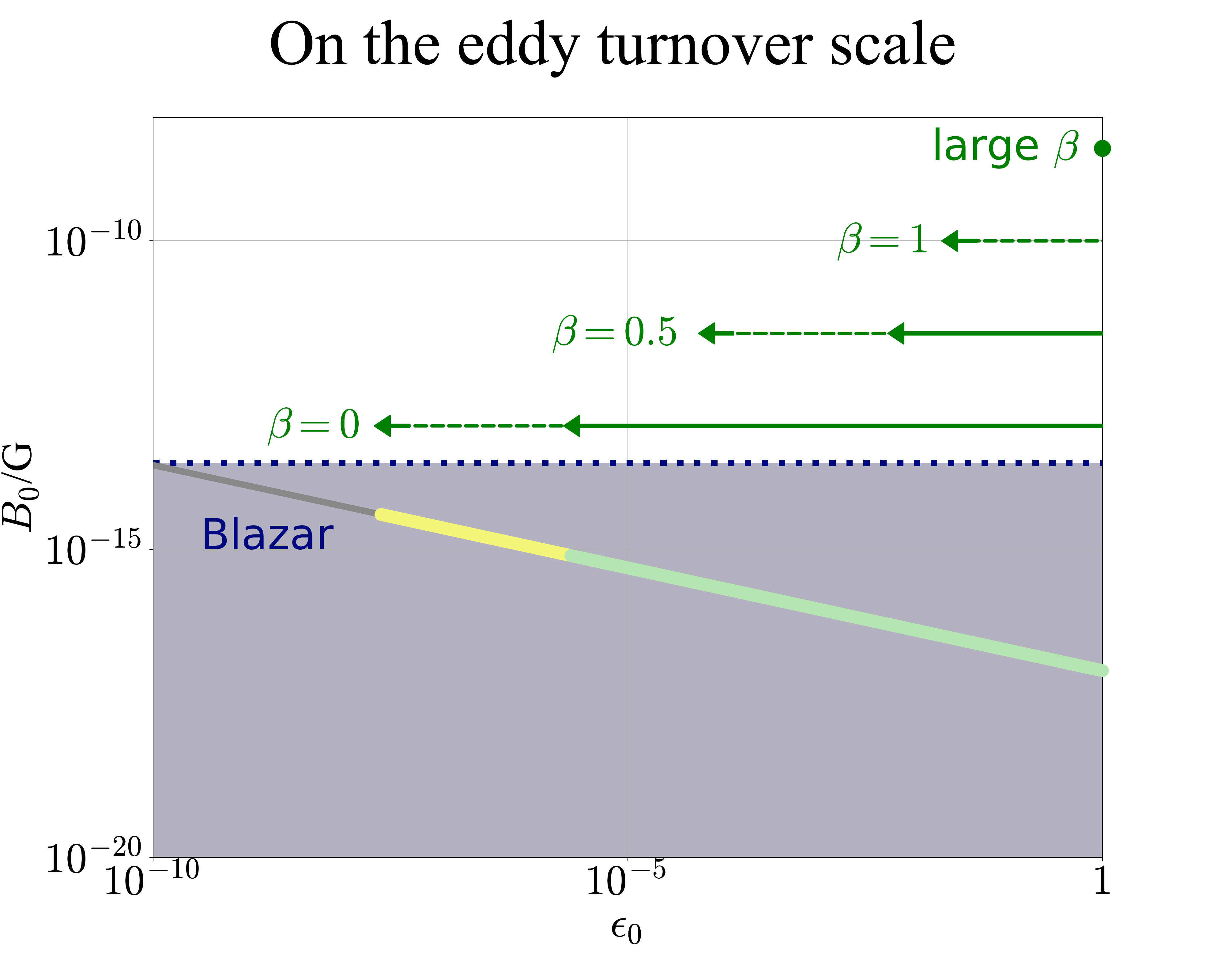}
\caption{\label{fig:2}
    The constraints on the present magnetic fields on the eddy turnover scale, Eqs. \eqref{eq:DeltaNow} and \eqref{eq:PowerNow}, are shown 
    as functions of the present helicity fraction.
    The thick green line is the allowed parameter region for the delta-function model,
    while the thick yellow line is allowed for the power-law model with $\alpha=0$, with the evolution parameter $\beta=0$.
    The navy dotted line shows the lower limit of the field strength extrapolated from the blazar observation \cite{Biteau:2018tmv}.
    The green arrows show the edges of the allowed regions for each $\beta$.
    For $\beta\gtrsim1$, only around $B_0=1\times10^{-17}\;$G with $\epsilon_0=1$ is allowed in both models.
    }
\end{center}
\end{figure}

Figure \ref{fig:2} shows the $\epsilon_0$ dependence of the IGMF strength on the MHD line Eq.~\eqref{eq:MHD}. 
Yellow and green lines represent the parameter region where the baryon isocurvature constraint is satisfied 
for $\beta=0$ in the case of the power-law spectrum  with an exponential cutoff and the delta-function spectrum, respectively. 
For larger $\beta$, the maximum field strength becomes smaller. 
In particular, for $\beta>1$, which would be often realized according to Ref.~\cite{Brandenburg:2017neh}, 
all the primordial magnetic fields that generate the correct BAU while satisfying the baryon isocurvature constraint
reaches at $B_0 \simeq 1\times 10^{-17} \;\mathrm{G}$. 
Comparing to the gray shaded region, where the IGMF cannot explain the blazar observation, 
we conclude that %it is hard to explain 
the primordial magnetic fields that generate the correct BAU alone
cannot explain the blazar observations regardless of its initial helicity or the shape of the spectrum.
Note that the spectrum of baryon isocurvature perturbations has more power on the smaller wavenumber modes than that of the energy density, 
and hence it is less sensitive to the behavior of the ultraviolet decay. 
Therefore, all the existing magnetogenesis mechanisms that explain the present BAU, 
such as the axion inflation~\cite{Anber:2015yca,Adshead:2016iae,Jimenez:2017cdr,Domcke:2018eki,Domcke:2019mnd,Barrie:2020kpt}, 
the chiral plasma instability~\cite{Kamada:2018tcs}, or the Affleck-Dine mechanism~\cite{Kamada:2019uxp}
are subject to this constraint, since each of these models produces
 helical hypermagnetic fields  at its own characteristic scale 
and they  evolve according to MHD so that their infrared spectrum is likely to become that of Batchelor spectrum.  One may wonder if 
it might be possible for some inflation models to generate partially helical magnetic fields at  Mpc scales today
while generating the net BAU, but such a possibility is now ruled out because of the excessive baryon isocurvature perturbation.

\section{Baryon isocurvature constraints on non-helical hypermagnetic fields}
\label{sec:Application}
Thus far we have discussed the baryon isocurvature constraint on the cosmological magnetic fields 
in the case they are responsible for the net BAU. 
Since not only the helical part but also the non-helical part sources the baryon isocurvature perturbation,
we can universally apply what we discussed in the previous sections to non-helical hypermagnetic fields
supposing the net BAU is generated by other mechanism, 
as long as magnetic fields existed before the EWSB.
In this section, we give the constraints on the non-helical hypermagnetic fields $\epsilon_\mathrm{fo} = \epsilon_0=0$ 
from the baryon isocurvature perturbation. 

The volume average of the baryon isocurvature pertuerbation in this case can be evaluated by setting $\epsilon_\mathrm{fo}=0$ and using the observed value of 
$\overline{\eta}_B$ in Eq. (\ref{eq:calGformula}) as
\begin{align}\label{eq:NonhelDelta}
    \overline{S^{2}_{B,\mathrm{BBN}}}= \left\{\begin{array}{ll}
         \dfrac{\mathcal{C}^2\xi_\mathrm{c,fo}^4 B_\mathrm{c,fo}^4}{2\overline{\eta}_B^2} D
         +\mathcal{O}\left(\left(\xi_\mathrm{c,fo}^2 D\right)^{2}\right) (<0.016),
         & \quad \text{for} \quad \xi_\mathrm{c,fo}^2 D\ll1,\hspace{15mm}  \\\\
         \dfrac{\mathcal{C}^2\xi_\mathrm{c,fo}^2 B_\mathrm{c,fo}^4}{3\overline{\eta}_B^2}
         +\mathcal{O}\left(\left(\xi_\mathrm{c,fo}^2  D\right)^{-1}\right) (<0.016),
         & \quad \text{for} \quad \xi_\mathrm{c,fo}^2  D\gg1
    \end{array}\right.
\end{align}
for the delta-function power spectrum and
\begin{align}\label{eq:NonhelPower}
    \overline{S^{2}_{B,\mathrm{BBN}}}= \left\{\begin{array}{l}
         \dfrac{\sqrt{\pi}2^{-\alpha}\Gamma\left(\frac{5}{2}+\alpha\right)\mathcal{C}^2\xi_\mathrm{c,fo}^5B^4_\mathrm{c,fo}}{4\Gamma\left(2+\frac{\alpha}{2}\right)^2\overline{\eta}_B^2}\left(\zeta_\alpha^2 D\right)^{\frac{3}{2}}
         +\mathcal{O}\left(\left[(\zeta_\alpha \xi_\mathrm{c,fo})^2 D\right]^{\frac{5}{2}}\right) (<0.016),\\
         \hspace{82mm}\text{for} \quad (\zeta_\alpha \xi_\mathrm{c,fo})^2 D\ll1,  \vspace{3mm}\\
         \dfrac{\Gamma\left(\frac{3\;+\;\alpha}{2}\right)\Gamma\left(\frac{5\;+\;\alpha}{2}\right)\mathcal{C}^2\xi_\mathrm{c,fo}^2 B^4_\mathrm{c,fo}}{3\Gamma\left(2+\frac{\alpha}{2}\right)\overline{\eta}_B^2}
         +\mathcal{O}\left(\left[(\zeta_\alpha \xi_\mathrm{c,fo})^2 D\right]^{-1}\right) (<0.016),\\
         \hspace{82mm}\text{for} \quad (\zeta_\alpha \xi_\mathrm{c,fo})^2  D\gg1
    \end{array}\right. 
\end{align}
for the power law spectrum with  an exponential cutoff.
In terms of physical quantities at the EWSB,
the magnetic field strength are then constrained as 
\begin{align}\label{eq:NonhelDeltaEWSB}
    B_\mathrm{p,fo} \lesssim 
    \left\{\begin{array}{ll}
    1 \times 10^{-7}  \ \mathrm{GeV}^2\left(\dfrac{\xi_{\rm p,fo}}{2 \times 10^{14}\;{\rm GeV}^{-1}}\right)^{-1},  
    & \quad \text{for} \quad \xi_\mathrm{p,fo}\ll \sqrt{\dfrac{2}{3}}k_\mathrm{d,fo}^{\mathrm{p} \ -1},\hspace{6mm}\\\\
    % to be more precise, 3.375
    % to be more precise, 14.83
    2 \times 10^{-7}   \ \mathrm{GeV}^{2} \left(\dfrac{\xi_{\rm p,fo}}{2 \times 10^{14}\;{\rm GeV}^{-1}}\right)^{-\frac{1}{2}},  
    & \quad \text{for} \quad \xi_\mathrm{p,fo}\gg \sqrt{\dfrac{2}{3}}k_\mathrm{d,fo}^{\mathrm{p} \ -1} 
    % to be more precise, 2.111
    % to be more precise, 14.83
    \end{array}\right. 
\end{align}
for the delta-function model, where $\sqrt{\dfrac{2}{3}}k_\mathrm{d,fo}^{\mathrm{p} \ -1} \simeq 2 \times 10^{14}\;{\rm GeV}^{-1}$. 
The constraints for the power-law model, Eqs. (\ref{eq:NonhelPower}), read
\begin{align}\label{eq:NonhelPowerEWSB}
\hspace{5mm}
    B_\mathrm{p,fo} \lesssim 
    \left\{\begin{array}{ll}
    1 \times 10^{-7} \;\mathrm{GeV}^2  \left(\dfrac{\xi_{\rm p,fo}}{2 \times 10^{14}\;{\rm GeV}^{-1}}\right)^{-\frac{5}{4}}\gamma_{1\alpha}^{\frac{1}{4}},& \; \text{for} \quad \xi_\mathrm{p,fo}\ll \sqrt{\dfrac{2}{3}}(\zeta_\alpha k_\mathrm{d,fo}^{\mathrm{p} })^{-1}  ,\\\\
    % to be more precise, 5.987
    % to be more precise, 14.83
    2 \times 10^{-7} \;\mathrm{GeV}^2  \left(\dfrac{\xi_{\rm p,fo}}{2 \times 10^{14}\;{\rm GeV}^{-1}}\right)^{-\frac{1}{2}}\gamma_{2\alpha}^{\frac{1}{4}},& \; \text{for} \quad \xi_\mathrm{p,fo}\gg \sqrt{\dfrac{2}{3}}(\zeta_\alpha k_\mathrm{d,fo}^{\mathrm{p}})^{-1} , 
    % to be more precise, 2.985
    % to be more precise, 14.83
    \end{array}\right. 
\end{align}
respectively, where we have defined $\gamma_{1\alpha} \equiv \dfrac{2^{\alpha}\Gamma(2+\frac{\alpha}{2})^5}{\Gamma(\frac{5\;+\;\alpha}{2})^3\Gamma(\frac{5}{2}+\alpha)}$ and $\gamma_{2\alpha} \equiv \dfrac{\Gamma(2+\frac{\alpha}{2})}{\Gamma(\frac{3\;+\;\alpha}{2})\Gamma(\frac{5\;+\;\alpha}{2})}$. 

\begin{figure}[htbp]
\begin{center}
    \includegraphics[width=0.48\textwidth]{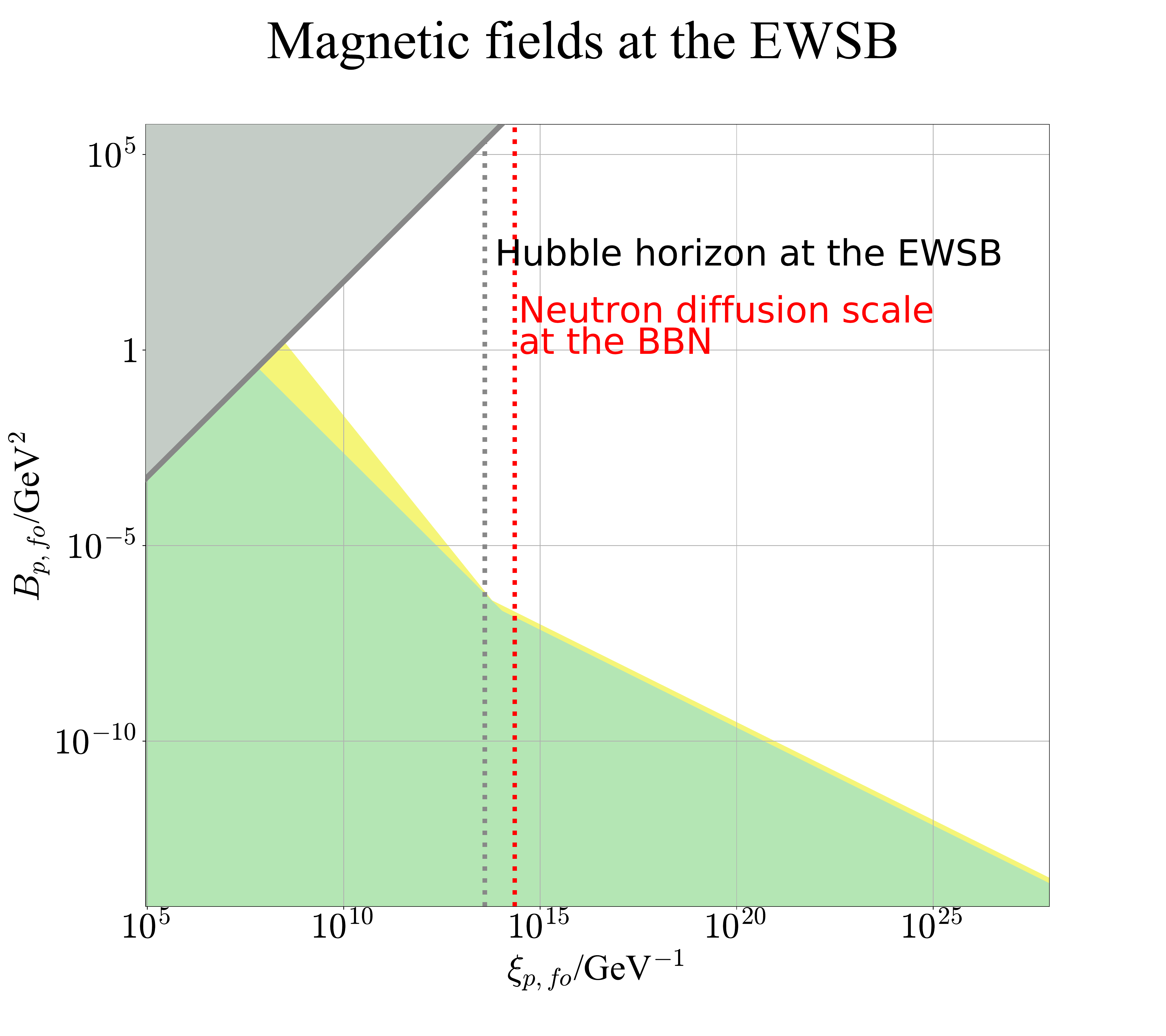}
    \includegraphics[width=0.48\textwidth]{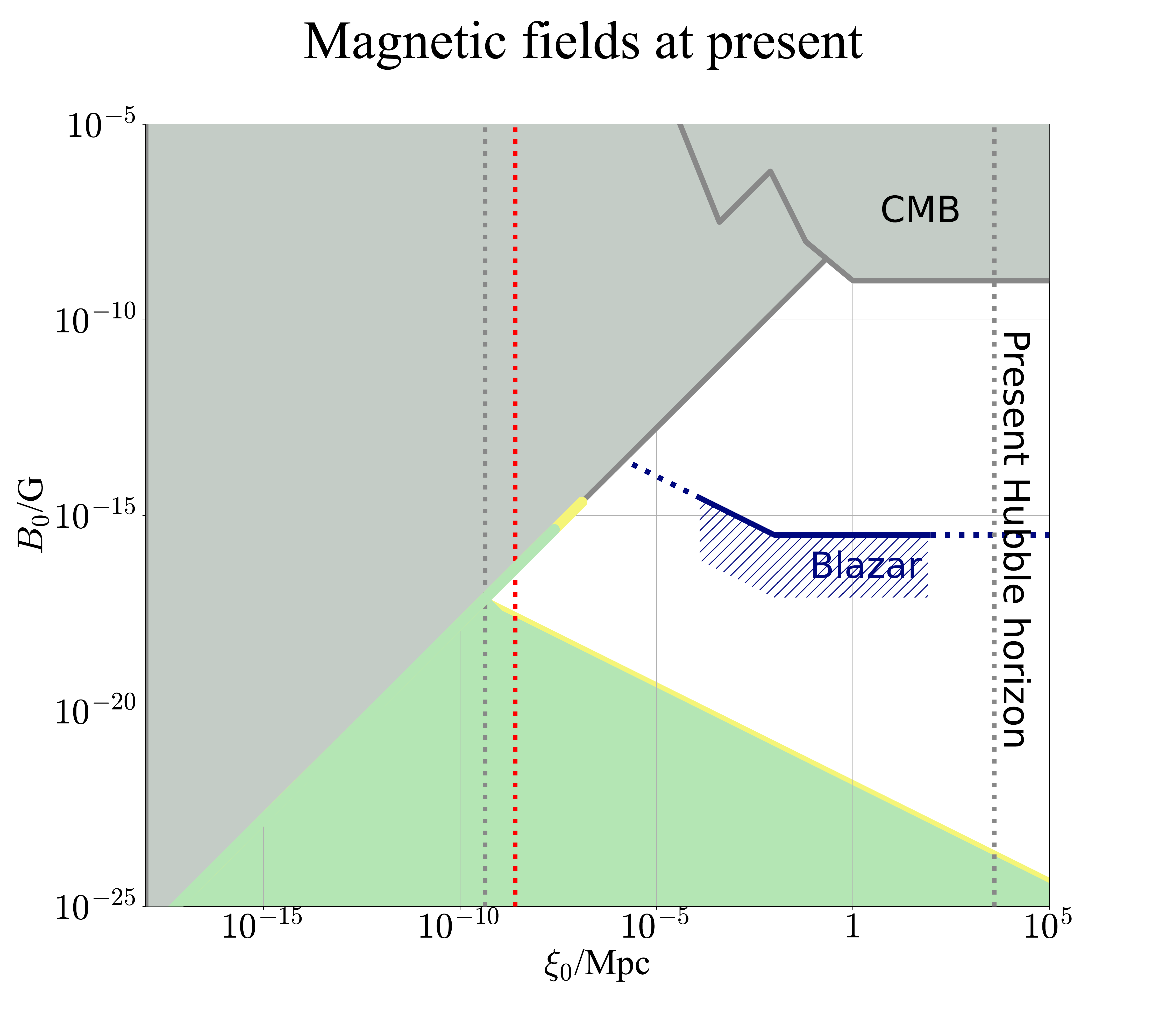}
\caption{\label{fig:5}
    Constraints on the magnetic fields at the EWSB, \eqref{eq:NonhelDeltaEWSB} and \eqref{eq:NonhelPowerEWSB} in terms of the physical quantities ({\it Left}) and 
    those on the present magnetic fields, Eqs. \eqref{eq:NonhelLong} and \eqref{eq:NonhelShort} ({\it Right}) are shown. 
    The navy line  in the right panel
    is the lower bounds of the IGMFs suggested by the blazar observations \cite{Biteau:2018tmv} with an extrapolation (dotted line). 
    In both panels,
    the gray shaded regions are the parameter spaces that are inconsistent with the MHD.
    The green shaded regions are the allowed parameter spaces for the delta-function model,
    while the yellow shaded regions are those for the power-law model with $\alpha=0$.
    }
\end{center}
\end{figure}

The parameter space satisfying these constraints is shown in the left panel of Fig.~\ref{fig:5} in green and yellow  for the 
delta-function model and the power-law model with an exponential cutoff, respectively.
Combining with the MHD turbulence condition at the EWSB (Eq.~\eqref{eq:eddyEWSB}), 
we find the upper bound of the magnetic field strength at the EWSB as
\begin{equation} \label{eq:strongnonhel}
    B_\mathrm{p,fo}^\mathrm{max} \simeq
    \left\{\begin{array}{ll} 
    0.4 \;\mathrm{GeV}^2, &\quad\text{for delta function},\\\\
    2 \;\mathrm{GeV}^2 \times \gamma_{1\alpha}^{-\frac{1}{9}}, &\quad\text{for power law},
    \end{array}\right. 
\end{equation}
respectively.  From the second lines of Eqs. \eqref{eq:NonhelDeltaEWSB} and \eqref{eq:NonhelPowerEWSB},
we can explicitly see that if the coherence length of the magnetic field at the EWSB is larger than the neutron diffusion scale,
the effect of neutron diffusion is so small that the 
volume average of the inhomogeneity of baryon distribution at the BBN is determined  only by $\xi_\mathrm{c,fo} B_\mathrm{c,fo}^2$ practically. 
This is because the amplitude of spatially-dependent component of the baryon asymmetry is still given by ${\cal C} |{\bm A}| \cdot |{\bm B}| \sim {\cal C} \xi_\mathrm{c,fo} B_\mathrm{c,fo}^2$ without the factor of the helicity fraction $\epsilon_\mathrm{fo}$, 
as can be seen in Eq.~\eqref{eq:BAUformula} regardless of the shape of the power spectrum.
On the other hand, in the case the coherence length of the magnetic fields is shorter than the neutron diffusion scale, 
the constraint depends on the shape of the spectrum.
See Appendix  \ref{sec:realspacecalG} for analysis in the position space.

Translating these conditions to the present magnetic fields,
we obtain the following constraints on the IGMF.
Large-scale magnetic fields compared with the eddy turnover scale at the recombination, Eq. \eqref{eq:MHD},
evolve adiabatically from EWSB until today, and hence 
the constraints for the magnetic fields at the EWSB~Eqs. \eqref{eq:NonhelDeltaEWSB} and \eqref{eq:NonhelPowerEWSB} is directly translated as
\begin{align}\label{eq:NonhelLong}
    B_0 \lesssim 
    \left\{\begin{array}{ll}
    3 \times 10^{-18} \;\mathrm{G} \left(\dfrac{\xi_0}{0.0025\;{\rm pc}}\right)^{-\frac{1}{2}}, &\quad\text{for delta function},\\\\
    % to be more precise, 1.536
    4 \times 10^{-18} \;\mathrm{G} \times \gamma_{2\alpha}^{\frac{1}{4}} \left(\dfrac{\xi_0}{0.0025\;{\rm pc}}\right)^{-\frac{1}{2}}, &\quad \text{for power law}. 
    % to be more precise, 2.172
    \end{array}\right. 
\end{align}
Note that the neutron diffusion scale accidentally coincides the scale where the constraint line intersects 
with the line for the MHD turbulence scale.

On the other hand, the  magnetic fields, which had a shorter coherence length originally, gradually start to evolve 
according to the cascade law and reach at the line that  is determined by the eddy turnover scale at the recombination 
(Eq.~\eqref{eq:MHD}). 
With a similar consideration as in the previous section, we find that 
the strongest magnetic fields at the present is obtained in the case with the strongest magnetic fields at the EWSB (Eq.~\eqref{eq:strongnonhel}) for the milder decay with smaller $\beta$ or 
in the case that magnetic fields enter the turbulence regime just before the recombination for the stronger decay with larger $\beta$. 
As a consequence, we obtain the upper bound of the magnetic field today as
\begin{align}\label{eq:NonhelShort}
    B_0 \lesssim 
    \left\{\begin{array}{ll}
    \text{max} \left[6\times 10^{-16}\;{\rm G}\times\left(\dfrac{\tau_{\rm rec}}{\tau_{\rm fo}}\right)^{\frac{1}{3}-\frac{\beta+1}{\beta+3}},\;6\times 10^{-18} \;\mathrm{G}\right], &\;\text{for delta function},\\\\
    % to be more precise, 1.611
    \text{max} \left[3\times10^{-15}\;{\rm G}
    \times\gamma_{1\alpha}^{-\frac{1}{9}} \times\left(\dfrac{\tau_{\rm rec}}{\tau_{\rm fo}}\right)^{\frac{1}{3}-\frac{\beta+1}{\beta+3}},\;8\times 10^{-18}\;\mathrm{G} \times \gamma_{2\alpha}^{\frac{1}{6}}\right], &\;\text{for power law}. 
    % to be more precise, 9.840
    \end{array}\right. 
\end{align}
The right panel of Fig.~\ref{fig:5} shows the parameter spaces that satisfy the baryon isocurvature constraint. 
Green and yellow shaded regions denote those for the delta-function model and power law with exponential cutoff model (for $\alpha=0$), 
respectively.  
The strongest magnetic fields are obtained at the MHD turbulence line (Eq.~\eqref{eq:MHD}), 
with the cascade law with $\beta=0$. 
Compared with the current observations,  once more, 
the upper bounds are below the blazar constraint, $B_0\lesssim3\times10^{-14}$ G, by a factor of $\mathcal{O}(10-100)$. 
On the contrary, the galaxy scales, $\xi_0 \sim 0.1-10$ kpc, the upper bound of the magnetic fields is roughly  $B_0\lesssim 10^{-20}$ G. 
This would be sufficient to be the seed magnetic fields for the magnetic fields in the galaxy and galaxy cluster~\cite{Davis:1999bt}.

Figures~\ref{fig:5} exhibit quite different allowed regions compared with  Fig.~\ref{fig:1} because the net baryon asymmetry is fixed 
independently of the properties of the magnetic fields. 
Since weaker magnetic fields that cannot generate the observed BAU is now allowed, 
viable parameter spaces are widely open. 
However, since the baryon isocurvature perturbation is mainly carried by the non-helical part of the hypermagnetic fields, 
the constraints do not change much for the stronger magnetic fields that can generate correct amount of the BAU, 
which lie on the line that is determined by the eddy turnover scale (Eq.~\eqref{eq:MHD}).

Thus far we have not discussed  the inflationary magnetogenesis that may have generated a relatively flat spectrum $\alpha < -4$
up to the Mpc scales today
with a comoving infrared cutoff $H_0<k_\mathrm{IR} < k_\mathrm{d} $ with $H_0$ being the present Hubble parameter. 
In this case, we can write the power spectrum as 
\begin{equation}
    S(k) = \frac{(B^\mathrm{IR}_\mathrm{c,fo})^2}{k_\mathrm{IR}^5} \left(\frac{k}{k_\mathrm{IR}}\right)^\alpha \exp\left[-\left(\frac{k}{k_\sigma}\right)^2\right]. 
\end{equation}
with $\epsilon_\mathrm{fo} = 0$. Here $B^\mathrm{IR}_\mathrm{c,fo}$ represents the magnetic field strength at the infrared cutoff.
In a similar way that is discussed in Eq.~\eqref{eq:IRdiverge}, we can evaluate the volume average of the baryon isocurvature perturbation 
as 
\begin{equation}
    \overline{S^{2}_{B,\mathrm{BBN}}} \sim \frac{\mathcal{C}^2 (B^\mathrm{IR}_\mathrm{c,fo})^4}{\overline{\eta}_B^2 k_\mathrm{IR}^2}, 
\end{equation}
which is almost equal to those for the delta-function like model or the power law with exponential cutoff model. 
Therefore, we can conclude that even in this case, the magnetic field strength at the IR cutoff should be tiny
so that the spectrum at the IR cut off comes below the constraint given in Eq.~\eqref{eq:NonhelLong}. 
We conclude that inflationary magnetogenesis with a relatively flat spectrum alone also cannot explain the blazar observations
if they are generated before the EWSB unless there are an amplification process.

\section{Conclusion and Discussion}
\label{sec:Discussion}
Baryogenesis from the hypermagnetic helicity decay is an attractive scenario since 
it does not require any new physical degrees of freedom beyond the SM in the mechanism itself. 
If this mechanism is responsible for the present BAU,  physics beyond the SM is required 
for the mechanism to produce the appropriate helical hypermagnetic fields. 

In this paper,
we have studied constraints on the mechanism from baryon isocurvature perturbations.
Since the baryon fluctuations are produced in response to the coherence length of the hypermagnetic fields, 
large scale hypermagnetic fields are disfavored otherwise it might produce the anti-matter region, 
or at least, spoils the success of the standard BBN.
One important essence is that non-helical part of the hypermagnetic fields contributes to the baryon isocurvature perturbation
while only their helical part contributes to the homogeneous part of the net baryon asymmetry.
Cosequently, baryon isocurvature perturbations  just with $\mathcal{O}(1)$~\% level on scales larger than the comoving neutron diffusion scale at BBN is ruled out 
because the deuteriums are overproduced due to the second order effect.

Our main results are summarized in Fig.~\ref{fig:1}. 
We find that the magnetic fields whose coherence length at the EWSB is larger than the comoving neutron diffusion scale at BBN 
are completely ruled out since generation of the correct amount of the BAU and the small enough baryon isocurvature 
perturbation cannot be satisfied simultaneously. In summary, 
for the primordial magnetic field to produce the BAU, 
the coherence length and the strength are constrained as $10^{6} \;{\rm GeV}^{-1} <\xi_{\rm p}<10^{15}\;{\rm GeV}^{-1}$ and $10^{-7}\; {\rm GeV}^2< B_{\rm p}< 10\;{\rm GeV}^2$ in physical quantity at the EWSB. 
Such magnetic fields evolve according to the MHD cascade processes and remain as the IGMFs. 
For the typical cascade parameter $\beta\simeq 2$, the present IGMF property is uniquely determined as 
$\xi_0 \simeq 10^{-3} \;\mathrm{pc}$ and $B_0 \simeq 10^{-17} \;\mathrm{G}$ with being maximally helical today, 
regardless of the shape of the magnetic field spectrum. 
Even allowing relatively unrealistic value of $\beta$ up to 0, we find the present IGMF must satisfy 
$10^{-3}\;{\rm pc} <\xi_{0}<1\;{\rm pc}$ and $10^{-17}\;{\rm G}< B_{0}<10^{-14}\;{\rm G}$, 
on the line Eq.~\eqref{eq:MHD}, which is determined by the eddy turnover scale.
It is found that the relationship among the parameters that describe the magnetic fields, that is, 
$B_0$, $\xi_0$), and $\epsilon_0$ is 
uniquely determined by the condition to generate the correct amount of BAU (Eq.~\eqref{eq:BAU_presentMF})
and to satisfy the MHD evolution (Eq.~\eqref{eq:MHD}) as $B_0 \propto \xi_0 \propto  \epsilon_0^{-\frac{1}{3}}$, 
 leaving only one degree of freedom to the IGMFs regardless of their spectrum or the evolution history. 
While this  would act as the important consistency relation to test this scenario,
the upper bound from the baryon isocurvature perturbation is (at most) marginally below the present lower bound of the IGMFs
suggested by the blazar observations~\cite{Neronov:1900zz,Tavecchio:2010mk,Ando:2010rb,Dolag:2010ni,Essey:2010nd,Taylor:2011bn,Takahashi:2013lba,Finke:2015ona,Biteau:2018tmv,AlvesBatista:2020oio}\footnote{Especially, if the constraint on the IGMF coherence length suggested in Ref.~\cite{AlvesBatista:2020oio} is correct,
it is quite difficult to satisfy both the baryon isocurvature constraints and IGMFs}.  

The fact that the non-helical part of the hypermagnetic fields contribute to the baryon isocurvature perturbation
enables us to constrain them at the EWSB, assuming that another mechanism is responsible for the BAU today. 
In this case, large-scale magnetic fields whose coherence length is larger than the comoving neutron diffusion scale at BBN
is permitted, but only tiny field satisfying  $B_0 \lesssim  10^{-18} \;\mathrm{G} (\xi_0/0.0025 \;\mathrm{pc})^{-1/2}$ today, 
which is well below the lower bound of the IGMFs suggested by the blazar observations. 
As for the IGMFs with a shorter wave length that have experienced the cascade process, 
the constraint is almost the same as the case when the baryogenesis from the hypermagnetic helicity decay 
is responsible for the present BAU. 
In summary, regardless of its helicity nature, hypermagnetic fields generated before the EWSB 
cannot be the origin of the IGMFs suggested by the blazar observations. 
We need another mechanism to generate or amplify the magnetic fields after EWSB. 
Note that the lower bound of the primordial magnetic fields required as a  source of the dynamo amplification 
of the galactic magnetic fields, $B_0\gtrsim 10^{-30}$ G at $\xi_0 \gtrsim 10^2$ pc~\cite{Davis:1999bt}, 
can be satisfied by the primordial hypermagnetic fields generated before the EWSB. 

There are several potential uncertainties in our analysis that could change our results. 
The first is that we have assumed that the symmetric and antisymmetric parts of the power spectrum, $S(q)$ and $A(q)$, are proportional to each other. 
However, the results are essentially unchanged even if this assumption is relaxed.
We see that Eq. \eqref{eq:calGformula} can be decomposed into contributions from the symmetric and antisymmetric parts as $\overline{S^{2}_{B,\mathrm{BBN}}}=\overline{\eta}_B^{-2}[\sum(\int SS{\rm-like\;terms})+\sum(\int AA{\rm-like\;terms})]$,
where the symmetric part should be the dominant contribution thanks to the realizability condition.
Therefore, by defining an effective helicity fraction, $\epsilon_{\rm eff}$, by $\epsilon_{\rm eff}^2\equiv\sum(\int AA{\rm-like\;terms})/\sum(\int SS{\rm-like\;terms})(\leq1)$,
the constraint is the same as that of \S \ref{sec:Concrete models}.

The second is that the temperature dependence of the weak mixing angle, which can change the 
efficiency factor of the baryogenesis, ${\cal C}$. 
In our analysis we adopted the value inspired by the one-loop analytic estimate~\cite{Kajantie:1996qd}, 
while the fitting formulae for the lattice simulation~\cite{DOnofrio:2015gop} adopted in Ref.~\cite{Kamada:2016cnb} 
can lead to a smaller value of ${\cal C}$. 
However, in a reasonable choice of the fitting formula, the constraints on the strength and coherence length magnetic fields
can be enhanced by a factor of ${\cal O}(10)$, but it is not enough for the upper bound from the baryon isocurvature 
constraints to reach at the lower bound of the IGMFs suggested by the blazar observations, 
except for the case when the magnetic fields have experienced the MHD cascade with milder decay $\beta=0$ (but see the 
next discussion).  
Since there are no accurate calculations on the efficiency factor ${\cal C}$ at the sphaleron freezeout $T\simeq 135$ GeV, 
we cannot rule out the possibility that it has a much smaller value in reality, but it would be unlikely. 
If some physics around the EWSB changes the weak mixing angle much more rapidly while keeping 
the sphaleron rate intact, it would be possible to realize a much smaller value of ${\cal C}$ so that 
small enough baryon isocurvature perturbations and the blazar observations might be simultaneously explained, 
which is beyond the scope of the present paper. 

Finally,  the MHD cascade evolution is treated in a relatively simplified way. 
We evaluate the eddy turnover scale (Eq.~\eqref{eq:MHD})  assuming the equipartition of the 
magnetic and fluid kinetic energy, which leads to $v \simeq v_\mathrm{A}$. 
If the eddy turnover scale is smaller, we might have a stronger IGMF today but with a smaller coherence length. 
In order for the remnant magnetic fields to reach at the lower bound of the IGMFs suggested by blazars, however,
we need the changes of the eddy turnover scale with more than a few orders of magnitude, 
which would be unlikely. 
Moreover, the upper bound of the IGMF strength in our analysis is obtained in the case with the cascade decay parameter $\beta=0$
at very tiny helicity. 
Evolution with $\beta=0$ is the case of the so-called inverse cascade with the maximally helical magnetic fields, 
which is guaranteed by the helicity conservation. 
We do not rule out such a possibility, but 
do not expect the evolution $\beta\simeq 0$ can be realized for the non-maximally helical magnetic fields, 
where there are no conserved quantity to support the evolution. 
In summary, although there still remain several uncertainties in our analysis, we conclude that they do not change our results
significantly.

What we have presented in this paper gives essential constraints on the model building of magnetogenesis. 
As we have shown that the strength of the IGMFs suggested by the blazar observations
is well above the baryon isocurvature constraints, any attempts to explain it by a primordial magnetic fields before the EWSB
is ruled out. For example, a first order phase transition in the physics beyond the SM at a higher energy scale than the EWSB 
does not work as a way to explain the origin of IGMFs. 
%On the inflationary magnetogenesis, we require at least %that the reheating completes after the EWSB, with %sufficiently large entropy production.
Hypermagnetic fields just to explain the BAU are still allowed, but the helicity fraction should not be too small, $\epsilon_\mathrm{fo} \gtrsim 10^{-8}$.
There has been a hope to test the scenario by IGMF observations, 
but we conclude that main part of the observationally suggested IGMFs should be generated or amplified after the EWSB, 
while magnetic fields responsible for the BAU is subdominant contribution of them. 
In order to test the scenario, we need to develop the way to remove such ``foreground'' IGMFs.\footnote{
Another option is to explain the blazar observation by some ways other than the IGMFs such as the beam plasma instability~\cite{Broderick:2011av}.}

\acknowledgments
The work of FU is supported by the Forefront Physics and Mathematics Program to Drive Transformation (FoPM). This work was partially supported by JSPS KAKENHI Grant-in-Aid for Scientific Research (C) JP19K03842 (KK),  Scientific Research (S) 20H05639(JY), Innovative Area 19H04610(KK), and 20H05248(JY).

\appendix
\section{Gauge choice}
\label{sec:Gauge choice}
Since the source term (\ref{eq:source}) is gauge dependent without taking a spatial average, 
we have to specify a gauge to carry out our calculation.
In the main text of the present paper,
we have imposed the Coulomb gauge condition,
$\boldsymbol{\nabla}\cdot\boldsymbol{A}_{\mathcal{A}}(\boldsymbol{x})=0.$
In this Appendix,
we see that we can determine the power spectrum of the vector potential, $\boldsymbol{A}_{\mathcal{A}}$,
as we do in \S \ref{sec:Definition},
by choosing the Coulomb gauge.

As for the magnetic field, $\boldsymbol{B}_{\mathcal{A}}$,
we can assume the homogeneity and the isotropy,
according to the cosmological principle.
Because of the homogeneity of the universe,
the two-point correlation function of the magnetic fields can be written as a function of the relative coordinate.
\eq{
    \langle B_{\mathcal{A}i}(\boldsymbol{x})B_{\mathcal{A}j}(\boldsymbol{x}+\boldsymbol{r})\rangle
    =
    \tilde{\mathcal{F}}^B_{ij}(\boldsymbol{r}).
    }
Transforming this equation into the wavenumber space,
we see that the power spectrum of the magnetic field is given by $\tilde{\mathcal{F}}_{ij}^B(\boldsymbol{k})$,
the Fourier transform of $\tilde{\mathcal{F}}_{ij}^B(\boldsymbol{x})$.
\eq{\label{eq:PowerSpectrumDef}
    \langle B^*_{\mathcal{A}i}(\boldsymbol{k})B_{\mathcal{A}j}(\boldsymbol{k}')\rangle
    =(2\pi)^3\delta^3(\boldsymbol{k}-\boldsymbol{k}')\tilde{\mathcal{F}}_{ij}^B(\boldsymbol{k}).
    }
Notice that the factor $\delta^3(\boldsymbol{k}-\boldsymbol{k}')$ appearing in the left hand side of the equation is a consequence of the homogeneity.
It immediately follows from Eq. \eqref{eq:PowerSpectrumDef} that $\tilde{\mathcal{F}}_{ji}^{B*}(\boldsymbol{k})=\tilde{\mathcal{F}}_{ij}^B(\boldsymbol{k})$.
Because of the isotropy and the divergenceless property of the magnetic field in addition to this,
the power spectrum of the magnetic field is characterized by two real-valued functions, $\tilde{S}^B(k)$ and $\tilde{A}^B(k)$ as 
\eq{
    \tilde{\mathcal{F}}^B_{ij}(\boldsymbol{k})
    =P_{ij}(\hat{\boldsymbol{k}})\tilde{S}^B(k)+ i\epsilon_{ijm}\hat{k}_m\tilde{A}^B(k),
    \quad P_{ij}(\hat{\boldsymbol{k}})
    =\delta_{ij}-\hat{k}_i\hat{k}_j.
    }

As for the vector potential, $\boldsymbol{A}_{\mathcal{A}},$
whether $\langle A^*_{\mathcal{A}i}(\boldsymbol{k})A_{\mathcal{A}j}(\boldsymbol{k}')\rangle$ has the factor $\delta^3(\boldsymbol{k}-\boldsymbol{k}')$ depends on the choice of gauge.
The reason why a gauge transformation breaks the condition equivalent to homogeneity is that values of the vector potential are not physical ones and that the homogeneity of the universe does not necessarily require the homogeneity of the vector potential field.
However,
with the Coulomb gauge condition,
the homogeneity is maintained,
and we can define the power spectrum of the vector potential,\footnote{We see this fact from $P_{ia}(\hat{\boldsymbol{k}})P_{jb}(\hat{\boldsymbol{k}}')\epsilon_{alr}\epsilon_{bms}\langle B^*_{\mathcal{A}l}(\boldsymbol{k})B_{\mathcal{A}m}(\boldsymbol{k}')\rangle=P_{ia}(\hat{\boldsymbol{k}})P_{jb}(\hat{\boldsymbol{k}}')k_rk'_s\langle A^*_{\mathcal{A}a}(\boldsymbol{k})A_{\mathcal{A}b}(\boldsymbol{k}')\rangle=k_rk'_s\langle A^*_{\mathcal{A}i}(\boldsymbol{k})A_{\mathcal{A}j}(\boldsymbol{k}')\rangle\propto\delta^3(\boldsymbol{k}-\boldsymbol{k}')$,
with the second equality valid under the Coulomb gauge condition, $P_{ij}(\hat{\boldsymbol{k}})A_{\mathcal{A}j}(\boldsymbol{k})=A_{\mathcal{A}i}(\boldsymbol{k})$.}
\eq{
    \langle A^*_{\mathcal{A}i}(\boldsymbol{k})A_{\mathcal{A}j}(\boldsymbol{k}')\rangle
    =:(2\pi)^3\delta^3(\boldsymbol{k}-\boldsymbol{k}')\tilde{\mathcal{F}}^A_{ij}(\boldsymbol{k}).
    }
Since $\boldsymbol{B}_{\mathcal{A}}=\boldsymbol{\nabla}\times\boldsymbol{A}_{\mathcal{A}},$
the power spectra, $\tilde{\mathcal{F}}_{ij}^B$ and $\tilde{\mathcal{F}}_{ij}^A$,
are related via
\eq{\label{eq:PowerSpectrumRelation}
    \tilde{\mathcal{F}}_{ij}^B(\boldsymbol{k})=\epsilon_{ilm}\epsilon_{jrs}k_lk_r\tilde{\mathcal{F}}_{ms}^A(\boldsymbol{k}).
    }
Similarly to the case of magnetic field,
$\tilde{\mathcal{F}}_{ji}^{A*}(\boldsymbol{k})=\tilde{\mathcal{F}}_{ij}^A(\boldsymbol{k})$,
the Coulomb gauge condition $k_i\tilde{\mathcal{F}}_{ij}^A(\boldsymbol{k})=0,$
and the isotropy of the magnetic field leaves only two real-valued functions, $\tilde{S}(k),\;\tilde{A}(k)$, as the degrees of freedom of the power spectrum of the vector potential,
\eq{
    \tilde{\mathcal{F}}^A_{ij}(\boldsymbol{k})
    =P_{ij}(\hat{\boldsymbol{k}})\tilde{S}(k)+ i\epsilon_{ijm}\hat{k}_m\tilde{A}(k).
    }
Using Eq. \eqref{eq:PowerSpectrumRelation}, $\tilde{S}^B,\;\tilde{A}^B$ and $\tilde{S},\;\tilde{A}$ are related via
\eq{
    \tilde{S}^B(k)=k^2\tilde{S}(k),\quad
    \tilde{A}^B(k)=k^2\tilde{A}(k).
    }
In addition to this,
as to the one-point function,
because of the Coulomb gauge condition,
\eq{
    \langle A_{\mathcal{A}i}(\boldsymbol{k})\rangle=0
    }
follows from the homogeneity of the magnetic field, $\langle B_{\mathcal{A}i}(\boldsymbol{k})\rangle=0$.\footnote{We see this from $-iP_{ia}(\hat{\boldsymbol{k}})\epsilon_{alr}\langle B_{\mathcal{A}l}(\boldsymbol{k})\rangle=P_{ia}(\hat{\boldsymbol{k}})k_r\langle A_{\mathcal{A}a}(\boldsymbol{k})\rangle=k_r\langle A_{\mathcal{A}i}(\boldsymbol{k})\rangle=0$,
with the second equality valid under the Coulomb gauge condition.}

The physical meaning of symmetric and antisymmetric parts of the power spectrum becomes clearer
if we use the popular notations
\eq{
    E_{\mathrm{M}}(k)=\frac{k^4}{2\pi^2}\tilde{S}(k),\quad
    H_{\mathrm{M}}(k)=\frac{k^3}{\pi^2}\tilde{A}(k).
    }
An integration of $E_{\mathrm{M}}(k)$ over $k$ gives the energy density of the fields, $\mathcal{E}=\int dkE_{\mathrm{M}}(k)$,
and an integration of $H_{\mathrm{M}}(k)$ the helicity density, $h=\int dkH_{\mathrm{M}}(k)$.
Therefore,
the symmetric part of the power spectrum determines the energy of the fields,
while the antisymmetric part determines the magnetic helicity of the fields.
One can explicitly see this property from Eqs. (\ref{eq:Strength}) and (\ref{eq:Helicity}) as well.

\section{Correlation functions in the position space}
\label{sec:realspacecalG}
In this appendix, 
we rederive the correlation function of the baryon isocurvature perturbation $\mathcal{G}$, defined in Eq. (\ref{eq:PtbCorr}),
in the real coordinate space.\footnote{Since we already have an expression in the wavenumber space, we can just Fourier transform it. However, here we work in the position space from the beginning. This reformulation will be helpful to intuitively understand how the baryon isocurvature perturbation is generated. See Ref. \cite{Giovannini:1997eg} for the earlier work about the procedure.}

\paragraph{Power spectrum in the position space}
We define three functions $F_1, F_2, F_3$, that characterize the power spectrum of the magnetic field in the position space, 
\eq{
    F_1 (R,t) \delta_{ij}+F_2 (R,t) R^iR^j\equiv
    \int\frac{d^3q}{(2\pi)^3}e^{+i\boldsymbol{q}\cdot\boldsymbol{R}}P_{ij}(\hat{\boldsymbol{q}})S(q),
    }
or equivalently
\begin{align}
    F_1(R)&=\frac{1}{2\pi^2}\int d(\ln q)\left[\left(-\frac{1}{R^3}+\frac{q^2}{R}\right)\sin(qR)+\frac{q}{R^2}\cos(qR)\right]S(q),\\
    F_2(R)&=\frac{1}{2\pi^2}\int d(\ln q)\left[\left(\frac{3}{R^5}-\frac{q^2}{R^3}\right)\sin(qR)-\frac{3q}{R^4}\cos(qR)\right]S(q),
\end{align}
and
\eq{\label{eq:F3:1}
    F_3({\bm R},t)\equiv
    \int\frac{d^3q}{(2\pi)^3}e^{+i\boldsymbol{q}\cdot\boldsymbol{R}}A(q).
    }
Here we have introduced a characteristic scale $k_\sigma$ that normalizes the dimensionful parameters, such as 
${\bm q} \equiv {\bm k}/k_\sigma$ and ${\bm R} \equiv k_\sigma {\bm x}$. 
$S(q)$ and $A(q)$ are also normalized by another parameter $B$, 
which characterizes the magnetic field strength so that the magnetic energy and helicity are evaluated as ${\cal E} \propto B^2$ 
and $h \propto B^2/k_\sigma$, respectively.
Then the power spectrum $\mathcal{F}_{ij}^A$ in the position space can be written in terms of these functions as
\eq{
    \mathcal{F}_{ij}^A({\bm R})=
    F_1 (R)\delta_{ij}+F_2(R)R^iR^j+\epsilon_{ijm}F_{3,m}({\bm R}),
    }
where $F_{3,m}({\bm R})\equiv\frac{\partial}{\partial R^{m}}F_3({\bm R})$.
$F_1$ and $F_2$ characterizes the symmetric part of the two-point correlation function of the magnetic field,
and $F_3$ does its antisymmetric part.
We shall note that $F_{3,m}$ has the following properties. 
By choosing the coordinate system so that the ${\bm R}$ lies in
the ${\bm e}_{3}$ direction, or $\boldsymbol{e}_3 \parallel\boldsymbol{R}$,  
for the coordinate ${\bm e}_{1,2}$, which are perpendicular to ${\bm e}_3$,
$F_{3,m}$ vanishes because the integrand of (\ref{eq:F3:1}) is odd with respect to $q_3$. 
Or we shall write  $F_{3,i}(\boldsymbol{R})=0$ for $i=1,2$.
On the other hand,
for the coordinate ${\bm e}_3$, $F_{3,3}$ gives finite value as 
\eq{
    F_{3,3}(\boldsymbol{R})
    =\frac{1}{2\pi^2R^2}\int dq\left[qR\cos(qR)-\sin(qR)\right]A(q)\equiv F_{3,3}(R,t), 
    }
as a function of the absolute value $R$. 
As long as $S(q)$ as well as $A(q)$ are localized at the scale $q\sim 1 \Leftrightarrow k \sim k_\sigma^{-1}$, 
$F_{1,2,3}(R)$ are also peaked at $R\sim 1 \Leftrightarrow r \sim k_\sigma^{-1}$. 
Note also that the net mean baryon asymmetry is given from Eq.~\eqref{eq:ProducedBAU} in terms of $F_{3,3}$ as
\begin{equation} \label{eq:Eta}
    {\overline \eta}_B = -6 {\cal C} k_\sigma^{-1}B^2F_{3,3}'(0)|_{T=T_\mathrm{fo}}, 
\end{equation}
where the prime denotes the derivative with respect to $R$. 
Particularly in the coordinate coordinate system with ${\bm e}_{i,j}$ with $i,j=1,2$  
being perpendicular to ${\bm R}$ whereas ${\bm e}_3$ is parallel to ${\bm R}$
so that ${\bm e}_1, {\bm e}_2$, and ${\bm e}_{3}$ construct the right-handed coordinate system (therefore $R^i=0$ for $i=1,2$ and $R^3=R$.), 
the Fourier transform of the power spectrum that contains the anti-symmetric part as
\begin{align}\label{eq:Formula1}
    \mathcal{F}^{A}_{ij}(\boldsymbol{R}) &= \int \frac{d^3 q}{(2\pi)^3} e^{+i {\bm q} {\bm R}} \mathcal{F}^{A}_{ij}(\boldsymbol{q}) 
    =F_1 (R)\delta_{ij}
    +\epsilon_{ij}F_3(R), 
\end{align} 
where $\epsilon_{ij}$ is the 2-dimensional Levi-Civita tensor with $\epsilon_{12}=1$. 
We shall note that other components of the magnetic field power spectrum are given as
\begin{align}
    \mathcal{F}^{A}_{i3}(\boldsymbol{R})= \mathcal{F}^{A}_{3 i}(\boldsymbol{R}) =0, \quad
    \mathcal{F}^{A}_{33}(\boldsymbol{R})= F_1(R)+F_2(R) R^2. \label{eq:Formula3}
\end{align}

\paragraph{Baryon isocurvature perturbation in the position space}
Let us now derive the expressions of the baryon isocurvature perturbations in terms of $F_i$s.
From Eq. \eqref{eq:BAUformula}, the two-point correlation function of the baryon isocurvature perturbation, $\mathcal{G}({\bm R})$, defined in Eq. \eqref{eq:PtbCorr}, is proportional to  a four-point function of the vector potential as
\eq{\label{eq:helicity_correlation}
    \langle \left(\boldsymbol{A}\cdot\nabla\times\boldsymbol{A}\right)(\boldsymbol{x})\left(\boldsymbol{A}\cdot\nabla\times\boldsymbol{A}\right)&(\boldsymbol{x}+\boldsymbol{r})\rangle\\
    =\epsilon_{ijk}\epsilon_{lmn} \lim_{{\bm s},{\bm t} \rightarrow 0}\pdiff{}{s^i}\pdiff{}{t^l}\langle &A_{k}(\boldsymbol{x}) A_{j}(\boldsymbol{x}+\boldsymbol{s})A_{n}(\boldsymbol{x}+\boldsymbol{r}) A_{m}(\boldsymbol{x}+\boldsymbol{r}+\boldsymbol{t})\rangle \\
    =\epsilon_{ijk}\epsilon_{lmn}  \lim_{{\bm s},{\bm t} \rightarrow 0}\pdiff{}{s^i}\pdiff{}{t^l}\Bigl[&\langle A_{k}(\boldsymbol{x}) A_{j}(\boldsymbol{x}+\boldsymbol{s})\rangle \langle A_{n}(\boldsymbol{x}+\boldsymbol{r}) A_{m}(\boldsymbol{x}+\boldsymbol{r}+\boldsymbol{t})\rangle \\
    &+\langle A_{k}(\boldsymbol{x}) A_{n}(\boldsymbol{x}+\boldsymbol{r}) \rangle \langle A_{j}(\boldsymbol{x}+\boldsymbol{s})  A_{m}(\boldsymbol{x}+\boldsymbol{r}+\boldsymbol{t})\rangle  \\
    &+\langle A_{k}(\boldsymbol{x}) A_{m}(\boldsymbol{x}+\boldsymbol{r}+\boldsymbol{t}) \rangle \langle A_{n}(\boldsymbol{x}+\boldsymbol{r})A_{j}(\boldsymbol{x}+\boldsymbol{s})  \rangle \Bigr] \\
    =\epsilon_{ijk}\epsilon_{lmn}  \lim_{{\bm s},{\bm t} \rightarrow 0}\pdiff{}{s^i}\pdiff{}{t^l}\Bigl[&\mathcal{F}^{A}_{kj}(k_{\sigma}\boldsymbol{s})\mathcal{F}^{A}_{nm}(k_{\sigma}\boldsymbol{t})+\mathcal{F}^{A}_{kn}(k_{\sigma}\boldsymbol{r})\mathcal{F}^{A}_{jm}(k_{\sigma}(\boldsymbol{r}-\boldsymbol{s}+\boldsymbol{t}))  \\
    &+ \mathcal{F}^{A}_{km}(k_{\sigma}(\boldsymbol{r}+\boldsymbol{t}))\mathcal{F}^{A}_{jn}(k_{\sigma}(\boldsymbol{r}-\boldsymbol{s})) \Bigr] \times k_{\sigma}^{-4}B^4, 
    }
In the second equality,
we decomposed the four-point function into pairs of two-point functions,
assuming the Gaussianity of the stochasticity.
In the last equality we have used the fact that the ensemble average $\langle A_i({\bm x}) A_j({\bm x}+{\bm r})\rangle $
is ${\bm x}$-independent, and hence $\langle A_i({\bm x}) A_j({\bm x}+{\bm r})\rangle= (1/V) \int d^3 x \langle A_i({\bm x}) A_j({\bm x}+{\bm r})\rangle = k_\sigma^{-2} B^2 \mathcal{F}^{A}_{ij} (k_{\sigma}{\bm r})$. 
Each contribution in Eq.~\eqref{eq:helicity_correlation} is calculated in terms of the functions $F_i$s as
\begin{subequations}
    \begin{align}
    &\epsilon_{ijk}\epsilon_{lmn}\lim_{{\bm s},{\bm t} \rightarrow 0}\pdiff{}{s^i}\pdiff{}{t^l}\left( \mathcal{F}^{A}_{kj}(k_{\sigma}\boldsymbol{s})\mathcal{F}^{A}_{nm}(k_{\sigma}\boldsymbol{t})\right) 
    =36k_{\sigma}^2F_{3,3}'^2(0),\label{eq:C:1}\\
    &\epsilon_{ijk}\epsilon_{lmn} \lim_{{\bm s},{\bm t} \rightarrow 0}\pdiff{}{s^i}\pdiff{}{t^l}\left(\mathcal{F}^{A}_{kn}(k_{\sigma}\boldsymbol{r})\mathcal{F}^{A}_{jm}(k_{\sigma}(\boldsymbol{r}-\boldsymbol{s}+\boldsymbol{t})) \right)\notag\\
    &\;=2k_{\sigma}^2\Biggl[-\frac{2}{R}F_1F_1'-F_1F_1''+3F_1F_2+RF_1F_2'-RF_1'F_2+R^2F_2^2 -F_{3,3}\left(F_{3,3}''+\frac{2 }{R}F_{3,3}'-\frac{2}{R^2}F_{3,3} \right)\Biggr],\label{eq:C:2}\\
    &\epsilon_{ijk}\epsilon_{lmn} \lim_{{\bm s},{\bm t} \rightarrow 0}\pdiff{}{s^i}\pdiff{}{t^l}\left(\mathcal{F}^{A}_{km}(k_{\sigma}(\boldsymbol{r}+\boldsymbol{t}))\mathcal{F}^{A}_{jn}(k_{\sigma}(\boldsymbol{r}-\boldsymbol{s})) \right) \notag \\
    &\;=2k_{\sigma}^2\Biggl(F_1'^2+R^2F_2^2-2RF_1'F_2+F'^2_3 + \frac{2}{R} F_{3,3} F'_{3,3} + \frac{3}{R^2} F_{3,3}^2\Biggr). \label{eq:C:3}
    \end{align}
\end{subequations}
Here we have omitted the arguments of the functions $F_i$ otherwise stated. 
Note that the final result is independent of the choice of the coordinate system.

The detailed calculations to obtain Eq. \eqref{eq:C:1}, \eqref{eq:C:1}, and \eqref{eq:C:3} are summarized in what follows.
We adopt the coordinate system such that the ${\bm R}$ lies in
the ${\bm e}_{3}$ direction and then can use Eqs. \eqref{eq:Formula1}, \eqref{eq:Formula3}.
First let us compute derivatives of $F_{3,3}$.
\begin{align}
    F_{3,311}(\boldsymbol{R})&= F_{3,322}(\boldsymbol{R}) = \frac{F_{3,3}(R)}{R}, \quad F_{3,333}(\boldsymbol{R}) = F_{3,3}'(R), \\
    \delta_{ij} F_{3,3ij} (\boldsymbol{R})&= - \int \frac{dq}{2 \pi^2} \frac{q^2}{R}\sin(qR) A(q) = F_{3,3}'(R)+\frac{2 F_{3,3}(R)}{R},  \\
    \delta_{ij} F_{3,3mij} (\boldsymbol{R})& = -\delta_{3m}\int \frac{dq}{2 \pi^2} \frac{q^2}{R^2} (qR \cos (qR)-\sin (qR)) A(q)\notag\\
    &= \delta_{3m} \left( F_{3,3}''(R)+\frac{2 F_{3,3}'(R)}{R} - \frac{2F_{3,3}(R)}{R^2}\right), 
\end{align}
where subscripts denote partial derivatives, {\it e.g.}, $F_{3,3ij}(\boldsymbol{R})\equiv\frac{\partial^2}{\partial R^j\partial R^i}R_{3,3}(R)$.
Now Eq.~\eqref{eq:C:1} is calculated as
\begin{align}
    \epsilon_{ijk}\epsilon_{lmn} \lim_{{\bm s},{\bm t} \rightarrow 0}\pdiff{}{s^i}\pdiff{}{t^l}\left( \mathcal{F}^{A}_{kj}(k_{\sigma}\boldsymbol{s})\mathcal{F}^{A}_{nm}(k_{\sigma}\boldsymbol{t})\right)  &= \left(\epsilon_{ijk} \lim_{{\bm s} \rightarrow 0} \frac{\partial}{\partial s^i} \mathcal{F}^{A}_{kj}(k_{\sigma}\boldsymbol{s}) \right)^2 \notag \\
    =\left(-2 \delta_{im} \lim_{{\bm s} \rightarrow 0} \frac{\partial}{\partial s^i} F_{3,3m}(k_\sigma {\bm s}) \right)^2 &= 36 k_\sigma^2 F_{3,3}'^2(0).
\end{align}
On the calculation of Eqs.~\eqref{eq:C:2} and \eqref{eq:C:3}, let us note that
\begin{equation} \label{eq:d5}
    \frac{\partial}{\partial R^j} F_i(R) = \frac{R^j}{R} F_i'(R), \quad \frac{\partial}{\partial R^j} \frac{\partial}{\partial R^m}F_i(R) = \frac{R^jR^m}{R^2} \left(F_i''(R) - \frac{F_i'}{R}\right)+\frac{F_i'}{R}\delta_{jm}, 
\end{equation}
for $i=1,2$, and hence
\begin{equation}
    \delta_{jm} \frac{\partial}{\partial R^j}\frac{\partial}{\partial R^m}F_i(R)  = F_i''(R) + \frac{2F_i'(R)}{R}.
\end{equation}
The first term in Eq.~\eqref{eq:d5} is non-vanishing only for $j=3$ so that
\begin{equation}
    \frac{\partial}{\partial R^j} F_i(R) =  F'_i(R) \delta_{j3},\quad \frac{\partial}{\partial R^3} \frac{\partial}{\partial R^3}  F_i(R) = F''_i(R). 
\end{equation}
By using these expressions, we reach the expression Eq.~\eqref{eq:C:2} as
\begin{align}
    \epsilon&_{ijk}\epsilon_{lmn}\lim_{{\bm s},{\bm t} \rightarrow 0}\pdiff{}{s^i}\pdiff{}{t^l} \left(\mathcal{F}^{A}_{kn}(k_{\sigma}\boldsymbol{r})\mathcal{F}^{A}_{jm}(k_{\sigma}(\boldsymbol{r}-\boldsymbol{s}+\boldsymbol{t})) \right) \notag \\
    &= \epsilon_{ijk}\epsilon_{lmn}k_\sigma^2 \biggl( F_1(R) \delta_{kn}+F_2(R) R_k R_n + \epsilon_{kns} F_{3,3s}(R) \biggr) \notag \\
    &\quad\times \lim_{{\bm S},{\bm T} \rightarrow 0} \frac{\partial}{\partial S^i}\frac{\partial}{\partial T^l} \biggl[F_1(|{\bm R}-{\bm S}+{\bm T}|)\delta_{jm} + F_2(|{\bm R}-{\bm S}+{\bm T}|) (R_j -S_j+T_j)(R_m-S_m+T_m)  \notag \\
    &\hspace{117.5mm}+ \epsilon_{jmt} F_{3,3t} ({\bm R}-{\bm S}+{\bm T}) \biggr] \notag \\
    &= 2 k_\sigma^2 \Biggl[-\frac{2}{R} F_1 F_1' - F_1F_1'' +3 F_1 F_2 +RF_1 F_2'-RF_1'F_2+R^2 F_2^2
    -F_{3,3}\left(F_{3,3}''+\frac{2}{R}F_{3,3}'-\frac{2}{R^2}F_{3,3} \right)\Biggr],
\end{align}
where we have defined as ${\bm S} \equiv k_\sigma {\bm s}$ and ${\bm T} \equiv k_\sigma {\bm t}$. 
In the same way, Eq.~\eqref{eq:C:3} can be derived as
\begin{align}
    \epsilon&_{ijk}\epsilon_{lmn} \lim_{{\bm s},{\bm t} \rightarrow 0}\pdiff{}{s^i}\pdiff{}{t^l}\left(\mathcal{F}^{A}_{km}(k_{\sigma}(\boldsymbol{r}+\boldsymbol{t}))\mathcal{F}^{A}_{jn}(k_{\sigma}(\boldsymbol{r}-\boldsymbol{s})) \right) \notag \\
    &= \epsilon_{ijk}\epsilon_{lmn}k_{\sigma}^2  \lim_{{\bm s},{\bm t} \rightarrow 0}\frac{\partial}{\partial T^l} \biggl[ F_1(|{\bm R}+{\bm T}|) \delta_{km} + F_2(|{\bm R}+{\bm T}|) (R_k+T_k) (R_m+T_m)+ \epsilon_{kmw} F_{3,3w} ({\bm R}+{\bm T})\biggr]\notag\\
    &\hspace{33mm}\times \frac{\partial}{\partial S^i} \biggl[ F_1(|{\bm R}-{\bm S}|) \delta_{jn} + F_2(|{\bm R}-{\bm S}|) (R_j-S_j) (R_n-S_n)+\epsilon_{jnz} F_{3,3z} ({\bm R}-{\bm S})\biggr] \notag \\
    &=2k_{\sigma}^2\biggl[F_1'^2+R^2F_2^2-2RF_1'F_2+F'^2_3 + \frac{2}{R} F_{3,3} F'_{3,3} + \frac{3}{R^2} F_{3,3}^2\biggr].
\end{align}

In summary, we obtain the two-point correlation function of the helicity in terms of the power spectrum as
\begin{align} \label{eq:helicity_corr_full}
    &\langle ({\bm A} \cdot {\bm \nabla} \times {\bm A})({\bm x})({\bm A} \cdot {\bm \nabla} \times {\bm A})({\bm x}+{\bm r}) \rangle - \langle ({\bm A} \cdot {\bm \nabla} \times {\bm A})({\bm x})  \rangle^2  \notag \\
    &\quad= 2 k_\sigma^{-2} B^4\Biggl[F_1'^2 -\frac{2}{R}F_1F_1'-F_1F_1''+3F_1F_2+RF_1F_2'-3RF_1'F_2+2R^2F_2^2 \notag \\
    &\hspace{73mm}+F_{3,3}'^2-F_{3,3}\left(F_{3,3}''-\frac{5}{R^2}F_{3,3} \right)\Biggr]. 
\end{align}
We can easily check that Eq.~\eqref{eq:helicity_corr_full} is consistent with the one given in Ref.~\cite{Giovannini:1997eg}.
Since in Ref.~\cite{Giovannini:1997eg} 
the magnetic fields are non-helical, $A(q)=0$, postulating that another mechanism is responsible for the BAU today, 
we have additional contributions from the helical part $A(q)$, which appear in terms of $F_{3,3}$.

Now we are ready to evaluate the baryon isocurvature perturbations in terms of the magnetic field power spectrum. 
From Eqs.~\eqref{eq:BAUformula}, \eqref{eq:Eta}, \eqref{eq:helicity_corr_full} 
we obtain
\begin{align}
    \mathcal{G}({\bm R})
    = \frac{{\cal C}^2}{\overline{\eta}_B^2} \left. \langle ({\bm A} \cdot {\bm \nabla} \times {\bm A})({\bm x})({\bm A} \cdot {\bm \nabla} \times {\bm A})({\bm x}+{\bm r}) \rangle - \langle ({\bm A} \cdot {\bm \nabla} \times {\bm A})({\bm x})  \rangle^2\right|_{T=T_\mathrm{fo}}
    =\frac{\tilde{G}(R)}{18 F'^2_{3,3}(0)},
\end{align}
with $R\equiv k_\sigma |{\bm r}|$ and 
\begin{align} \label{eq:Eta_Corr}
    {\tilde G}(R)\equiv &F_1'^2(R) -\frac{2}{R}F_1(R)F_1'(R)-F_1(R)F_1''(R)+3F_1(R)F_2(R)+RF_1(R)F_2'(R)\notag \\& -3RF_1'(R)F_2(R)+2R^2F_2^2(R) 
    +F_{3,3}'^2(R)-F_{3,3}(R)\left(F_{3,3}''(R)-\frac{5}{R^2}F_{3,3}(R) \right).
\end{align}

\paragraph{Delta function spectrum}
In the case of the delta-function model, $S(q)$ and $A(q)$,
\begin{equation}
    S(q)=\delta(q-1), \quad A(q) = \epsilon \delta(q-1), 
\end{equation}
$F_i$s become
\begin{subequations}
    \begin{align}
    F_1(R)
    &=\frac{1}{2\pi^2}\left[\left(\frac{1}{R}-\frac{1}{R^3}\right)\sin R+\frac{1}{R^2}\cos R\right],\\
    F_2(R)
    &=\frac{1}{2\pi^2}\left[\left(-\frac{1}{R^3}+\frac{3}{R^5}\right)\sin R-\frac{3}{R^4}\cos R\right],\\
    F_{3,3}(R)&=\frac{\epsilon}{2\pi^2}\left[-\frac{1}{R^2}\sin R+\frac{1}{R}\cos R\right].
    \end{align}
\end{subequations}
Then inserting them into Eq. \eqref{eq:Eta_Corr},
we can compute the correlator of baryon isocurvature perturbations as
\eq{\label{eq:GDelta}
    \mathcal{G}(\boldsymbol{R})
    &=\frac{1}{4}\left(\frac{1}{\epsilon^2}+1\right)\left[\frac{2}{R^2}+\frac{2}{R^4}+\frac{3}{R^6}-\frac{6}{R^5}\sin 2R+\left(\frac{4}{R^4}-\frac{3}{R^6}\right)\cos2R\right].
    }

\paragraph{Power-law-like spectrum}
In the case of the power law with exponential cutoff model, $S(q)$ and $A(q)$, 
\begin{equation}
    S(q) = q^\alpha \exp(-q^2), \quad A(q) = \epsilon q^\alpha \exp(-q^2),
\end{equation}
$F_i$s become
\begin{subequations}
    \begin{align}
    F_1(R)
    &=\frac{\Gamma(\frac{3\;+\;\alpha}{2})}{60\pi^2}\left[10F(\frac{3+\alpha}{2},\frac{5}{2},-\frac{R^2}{4})-\frac{3+\alpha}{2}F(\frac{5+\alpha}{2},\frac{7}{2},-\frac{R^2}{4})R^2\right],\\
    F_2(R)
    &=\frac{\Gamma(\frac{5\;+\;\alpha}{2})}{60\pi^2}F(\frac{5+\alpha}{2},\frac{7}{2},-\frac{R^2}{4}),\\
    F_{3,3}(R)
    &=\frac{\Gamma(1+\frac{\alpha}{2})\epsilon}{4\pi^2}\left[F(1+\frac{\alpha}{2},\frac{1}{2},-\frac{R^2}{4})-F(1+\frac{\alpha}{2},\frac{3}{2},-\frac{R^2}{4})\right]\frac{1}{R},
    \end{align}
\end{subequations}
where $F(a, c, z)$ is the confluent hypergeometric function.
Then inserting them into Eq. \eqref{eq:Eta_Corr},
we can compute the correlator of baryon isocurvature perturbations.
The asymptotic behavior in the limit $R\to0$, $\mathcal{G}(\boldsymbol{R})$ is
\eq{\label{eq:GAsymptotic}
    \mathcal{G}(\boldsymbol{R})
    =\frac{1}{3}\left[\frac{\Gamma\left(\frac{3\;+\;\alpha}{2}\right)\Gamma\left(\frac{5\;+\;\alpha}{2}\right)}{\Gamma\left(2+\frac{\alpha}{2}\right)^2\epsilon^2}+1\right]-\frac{1}{18}\left[\frac{\Gamma\left(\frac{3\;+\;\alpha}{2}\right)\Gamma\left(\frac{7\;+\;\alpha}{2}\right)}{\Gamma\left(2+\frac{\alpha}{2}\right)^2\epsilon^2}+2+\frac{\alpha}{2}\right]R^2+\mathcal{O}(R^4),
    }
while in the opposite limit $R\to\infty$,
\eq{\label{eq:GPower}
    \mathcal{G}(\boldsymbol{R})
    =\frac{\Gamma(3+\alpha)^2}{\Gamma(2+\frac{\alpha}{2})^2}\left[\frac{1}{\epsilon^2}(3+\alpha)(5+\alpha)\frac{-1+\cos(\pi\alpha)}{\alpha(2+\alpha)^2}+(2-\alpha)\frac{1+\cos(\pi\alpha)}{(1+\alpha)^2}\right]R^{-8-2\alpha}\;&\\
    +\mathcal{O}(R^{-10-2\alpha})&.
    }

\begin{figure}
\begin{center}
    \includegraphics[angle = -90, width=0.7\textwidth]{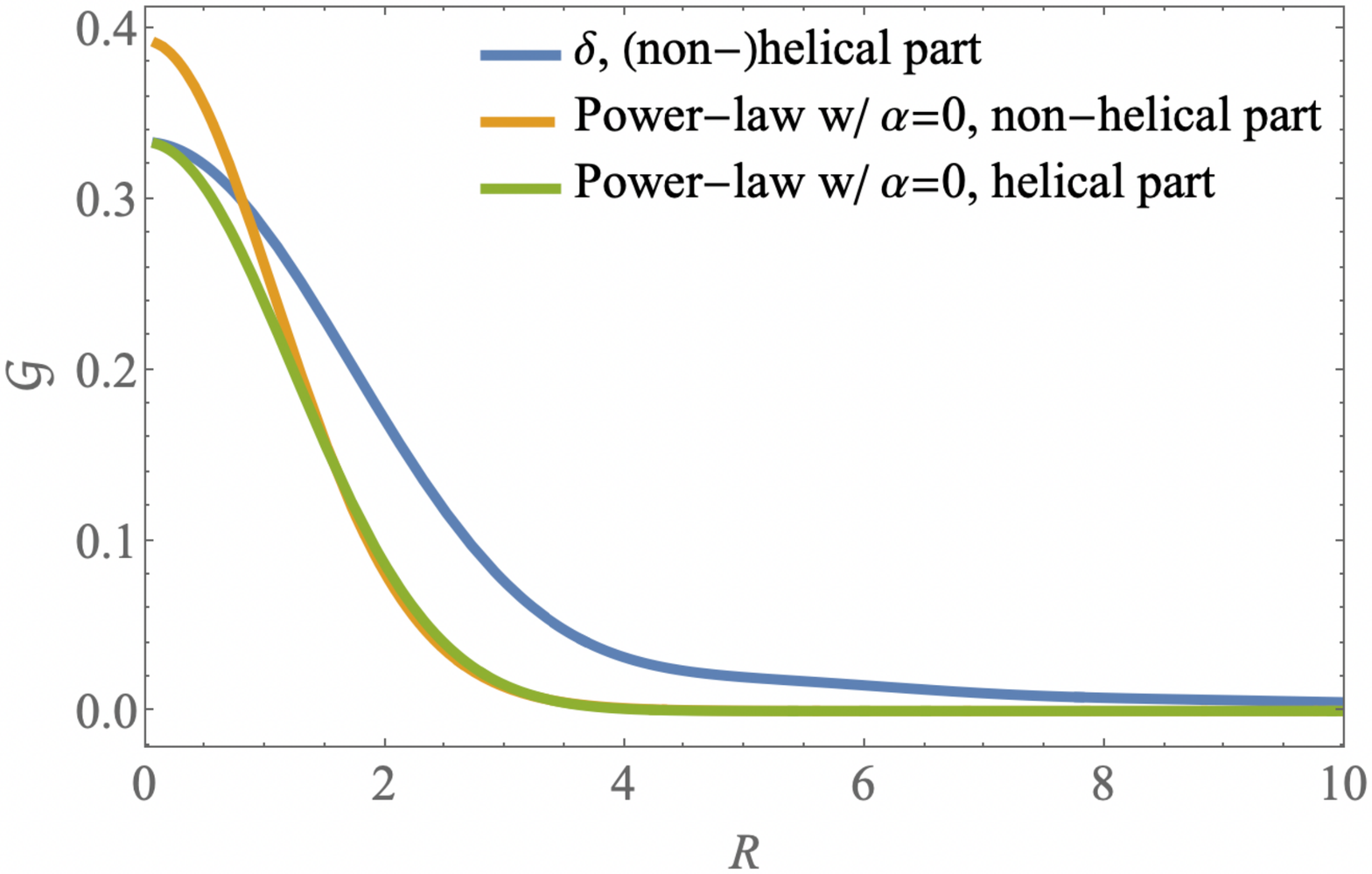}
\caption{\label{fig:4}
    Comparison of the shapes of $\mathcal{G}(R)$ with the delta-function model and the power-law model with $\alpha=0$ is shown. $\mathcal{G}(R)$ has terms proportional to $1/\epsilon^2$ originated from the non-helical part of the spectrum, and terms independent of $\epsilon$ originated from the helical part. In the delta-function model, the correlation function
    is exactly proportional to $1+1/\epsilon^2$.}
\end{center}
\end{figure}

\paragraph{Comparison of models}
Finally,
let us compare the behaviors of the correlation function of the baryon isocurvature perturbation
for the delta-function model and the power-law model in the position space, 
which enables us to understand those in the wavenumber space presented in the main text.
Since spatially averaged baryon asymmetry (\ref{eq:Eta}) is independent of the choice of $S(q)$,
no difference between the delta-function model and the power-law model appears in the constraint from the  overproduction of baryon asymmetry.
However,
as to the constraint of deuterium overproduction,
Fig.~\ref{fig:2} shows a trend of weaker constraint for the power-law model compared to the delta-function model.
The origin of this trend is not so trivial,
because the four-point function is not linear in the power spectrum and thus Fourier modes are not independent of each other.
An intuitive interpretation is can be given as follows. 
The delta-function model has a perfect periodicity and thus non-vanishing long-range correlation in the position space.
If we compare asymptotic behavior of $\mathcal{G}(R)$ in $R\to \infty$ limit,
for the delta-function model,
from (\ref{eq:GDelta}), we find 
$\mathcal{G}\sim R^{-2}$ when $R\to\infty$.
On the other hand for the power-law model,
from (\ref{eq:GPower}), we find
$\mathcal{G}\sim R^{-8-2\alpha}$ when $R\to\infty$.
Since our treatment is valid only for $\alpha>-3$,
the tail in the large $R$ limit in the power-law model is more suppressed than that in the delta-function model, 
as can also be seen in Fig.~\ref{fig:4}. 
Since the neutron diffusion effect erases the small scale correlation, 
constraints on the models with stronger large-scale tail, that is, the delta-function model, are severer than that of the power-law model, 
which explains the tendency in Fig.~\ref{fig:2}.

\end{document}